\documentclass[12pt]{article}
\usepackage{amssymb}

\newtheorem{theorem}{Theorem}[section]
\newtheorem{corollary}{Corollary}[section]
\newtheorem{lemma}{Lemma}[section]

\begin{document}

\author{C. C. Ciob\^{\i }rc\u {a}\thanks{%
e-mail address: ciobarca@central.ucv.ro}, E. M. Cioroianu\thanks{%
e-mail address: manache@central.ucv.ro}, S. O. Saliu\thanks{%
e-mail address: osaliu@central.ucv.ro} \\
Faculty of Physics, University of Craiova\\
13 A. I. Cuza Str., Craiova 200585, Romania}
\title{Cohomological BRST aspects of the massless tensor field with the mixed
symmetry $\left( k,k\right) $}
\maketitle

\begin{abstract}
The main BRST cohomological properties of a free, massless tensor field that
transforms in an irreducible representation of $GL\left( D,\mathbb{R}\right) 
$, corresponding to a rectangular, two-column Young diagram with $k>2$ rows
are studied in detail. In particular, it is shown that any non-trivial
co-cycle from the local BRST cohomology group $H\left( s|d\right) $ can be
taken to stop either at antighost number $\left( k+1\right) $ or $k$, its
last component belonging to the cohomology of the exterior longitudinal
derivative $H\left( \gamma \right) $ and containing non-trivial elements
from the (invariant) characteristic cohomology $H^{\mathrm{inv}}\left(
\delta |d\right) $.

\end{abstract}

\section{Introduction}

An interesting class of field theories is represented by tensor fields in
``exotic'' representations of the Lorentz group, characterized by a mixed
Young symmetry type~\cite{curt,aul,labast,burd,zinov1}, which are known to
appear in superstring theories, supergravities or supersymmetric high spin
theories. This type of models became of special interest lately due to the
many desirable featured exhibited, like the dual formulation of field
theories of spin two or higher~\cite
{dualsp1,dualsp2,dualsp2a,dualsp3,dualsp4,dualsp5}, the impossibility of
consistent cross-interactions in the dual formulation of linearized gravity~%
\cite{lingr} or a Lagrangian first-order approach~\cite{zinov2,zinov3} to
some classes of massless or partially massive mixed symmetry-type tensor
gauge fields, suggestively resembling to the tetrad formalism of General
Relativity. A basic problem involving mixed symmetry-type tensor fields is
the approach to their local BRST cohomology, since it is helpful at solving
many Lagrangian and Hamiltonian aspects, like, for instance the
determination of their consistent interactions~\cite{def} with higher-spin
gauge theories~\cite{epjc,high1,high2,high3,high4,noijhep,hepth04,boulanger}%
. The present paper proposes the investigation of the basic cohomological
ingredients involved in the structure of the co-cycles from the local BRST
cohomology for a free, massless tensor gauge field $t_{\mu _{1}\cdots \mu
_{k}|\nu _{1}\cdots \nu _{k}}$ that transforms in an irreducible
representation of $GL\left( D,\mathbb{R}\right) $, corresponding to a
rectangular, two-column Young diagram with $k>2$ rows.

In view of this, we firstly give the Lagrangian formulation of such a mixed
symmetry tensor field from the general principle of gauge invariance and
then systematically analyze this formulation in terms of the generalized
differential complex~\cite{genpoinc} $\Omega _{2}\left( \mathcal{M}\right) $
of tensor fields with mixed symmetries corresponding to a maximal sequence
of Young diagrams with two columns, defined on a pseudo-Riemannian manifold $%
\mathcal{M}$ of dimension $D$. Secondly, we compute the associated free
antifield-BRST symmetry $s$, which is found to split as the sum between the
Koszul-Tate differential and the exterior longitudinal derivative only, $%
s=\delta +\gamma $. Thirdly, we pass to the cohomological approach to this
model and prove the following results:

\begin{itemize}
\item  the cohomology of the exterior longitudinal derivative $H\left(
\gamma \right) $ is non-trivial only in pure ghost numbers of the type $kl$,
with $l$ any non-negative integer;

\item  both the cohomologies of the exterior spacetime differential $d$ in
the space of invariant polynomials and in $H\left( \gamma \right) $ are
trivial in strictly positive antighost number and in form degree strictly
less than $D$;

\item  there is no non-trivial descent for $H\left( \gamma |d\right) $ in
strictly positive antighost number;

\item  the invariant characteristic cohomology $H^{\mathrm{inv}}\left(
\delta |d\right) $ is trivial in antighost numbers strictly greater than $%
\left( k+1\right) $;

\item  any co-cycle from the local BRST cohomology $H\left( s|d\right) $ of
definite ghost number and in form degree $D$ can be made to stop at a
maximum value of the antighost number equal to either $k$ or $\left(
k+1\right) $ by trivial redefinitions only;

\item  the non-trivial piece of highest antighost number from any such
co-cycle can always be taken to belong to $H\left( \gamma \right) $, with
some coefficients that are non-trivial elements from $H^{\mathrm{inv}}\left(
\delta |d\right) $.
\end{itemize}

The results contained in this paper can be used at the determination of the
consistent couplings between the free, massless tensor field with the mixed
symmetry $\left( k,k\right) $ and other matter and gauge fields.

\section{Lagrangian formulation from the principle of gauge invariance\label%
{2}}

We consider a tensor field $t_{\mu _{1}\cdots \mu _{k}|\nu _{1}\cdots \nu
_{k}}$ that transforms in an irreducible representation of $GL\left( D,%
\mathbb{R}\right) $, corresponding to a rectangular, two-column Young
diagram with $k>2$ rows 
\begin{equation}
t_{\mu _{1}\cdots \mu _{k}|\nu _{1}\cdots \nu _{k}}= 
\begin{array}{ll}
\mu _{1} & \nu _{1} \\ 
\vdots & \vdots \\ 
\mu _{k} & \nu _{k}
\end{array}
,  \label{k1}
\end{equation}
or, in a shortened version, a tensor field with the mixed symmetry $\left(
k,k\right) $. This means that $t_{\mu _{1}\cdots \mu _{k}|\nu _{1}\cdots \nu
_{k}}$ is separately antisymmetric in the first and respectively last $k$
indices, is symmetric under the inter-change between the two sets of indices 
\begin{equation}
t_{\mu _{1}\cdots \mu _{k}|\nu _{1}\cdots \nu _{k}}=t_{\nu _{1}\cdots \nu
_{k}|\mu _{1}\cdots \mu _{k}},  \label{k2}
\end{equation}
and satisfies the (algebraic) Bianchi I identity 
\begin{equation}
t_{\left[ \mu _{1}\cdots \mu _{k}|\nu _{1}\right] \nu _{2}\cdots \nu
_{k}}\equiv 0.  \label{k3}
\end{equation}
Here and in the sequel the symbol $\left[ \mu \cdots \nu \right] $ signifies
the operation of complete antisymmetrization with respect to the indices
between brackets, defined such as to include only the distinct terms for a
tensor with given antisymmetry properties. For instance, the left-hand side
of (\ref{k3}) contains precisely $\left( k+1\right) $ terms 
\begin{eqnarray}
t_{\left[ \mu _{1}\cdots \mu _{k}|\nu _{1}\right] \nu _{2}\cdots \nu _{k}}
&\equiv &t_{\mu _{1}\cdots \mu _{k}|\nu _{1}\nu _{2}\cdots \nu _{k}}+\left(
-\right) ^{k}t_{\mu _{2}\cdots \mu _{k}\nu _{1}|\mu _{1}\nu _{2}\cdots \nu
_{k}}  \nonumber \\
&&+t_{\mu _{3}\cdots \mu _{k}\nu _{1}\mu _{1}|\mu _{2}\nu _{2}\cdots \nu
_{k}}+\cdots +\left( -\right) ^{k}t_{\nu _{1}\mu _{1}\cdots \mu _{k-1}|\mu
_{k}\nu _{2}\cdots \nu _{k}}.  \label{k4}
\end{eqnarray}
Assume that this tensor field is defined on a pseudo-Riemannian manifold $%
\mathcal{M}$ of dimension $D$, like, for instance, a Minkowski-flat
spacetime of dimension $D$, endowed with a metric tensor of `mostly plus'
signature $\sigma _{\mu \nu }=\sigma ^{\mu \nu }=\left( -+\cdots +\right) $.
The various traces of this tensor field, to be denoted by $t_{\mu _{1}\cdots
\mu _{k-m}|\nu _{1}\nu _{2}\cdots \nu _{k-m}}$, are defined by 
\begin{equation}
t_{\mu _{m+1}\cdots \mu _{k}|\nu _{m+1}\nu _{2}\cdots \nu _{k}}=\sigma ^{\mu
_{1}\nu _{1}}\cdots \sigma ^{\mu _{m}\nu _{m}}t_{\mu _{1}\cdots \mu _{k}|\nu
_{1}\nu _{2}\cdots \nu _{k}},\;m=\overline{1,k},  \label{k5}
\end{equation}
plus the conventions 
\begin{equation}
f_{\mu _{m+1}\mu _{m}}\equiv f\;(\mathrm{scalar}),\;f_{\mu _{m}\mu
_{m}}\equiv f_{\mu _{m}}\;(\mathrm{vector}).  \label{conv}
\end{equation}
Obviously, each type of trace transforms in an irreducible representation of 
$GL\left( D,\mathbb{R}\right) $, corresponding to a rectangular, two-column
Young diagram with $\left( k-m\right) $ rows 
\begin{equation}
t_{\mu _{m+1}\cdots \mu _{k}|\nu _{m+1}\cdots \nu _{k}}= 
\begin{array}{ll}
\mu _{m+1} & \nu _{m+1} \\ 
\vdots & \vdots \\ 
\mu _{k} & \nu _{k}
\end{array}
,  \label{k6}
\end{equation}
i.e., it is separately antisymmetric in the first and respectively last $%
\left( k-m\right) $ indices, is symmetric under the inter-change between the
two sets of indices and satisfies the identity 
\begin{equation}
t_{\left[ \mu _{m+1}\cdots \mu _{k}|\nu _{m+1}\right] \nu _{m+2}\cdots \nu
_{k}}\equiv 0,  \label{k7}
\end{equation}
which results from (\ref{k3}) by some appropriate contractions.

We are interested in the Lagrangian description of a single, free, massless
tensor field with this type of mixed symmetry, which is known to describe
exotic spin-two particles for $k\geq 2$. For $k=1$ we obtain nothing but the
spin-two field in the linearized limit of General Relativity, known as the
Pauli-Fierz theory \cite{pf}, while for $k=2$ we recover the free, massless
tensor field with the mixed symmetry of the Riemann tensor \cite{zinov1,epjc}%
. \textit{The construction of the Lagrangian action for such a tensor field
relies on the general principle of gauge invariance, combined with the
requirements of locality, Lorentz covariance, Poincar\'{e} invariance, zero
mass and the natural assumptions that the field equations are linear in the
field, second-order derivative and do not break the PT invariance.} In view
of all these, a natural point to start with is to stipulate the
(infinitesimal) gauge invariance of the action such that to recover the
linearized limit of diffeomorphisms for $k=1$ and the gauge symmetry \cite
{zinov1,epjc} of the free, massless tensor field with the mixed symmetry of
the Riemann tensor for $k=2$. The simplest way to achieve this is to ask
that the Lagrangian action $S^{\mathrm{L}}\left[ t_{\mu _{1}\cdots \mu
_{k}|\nu _{1}\cdots \nu _{k}}\right] $ is invariant under the
(infinitesimal) gauge transformations 
\begin{equation}
\delta _{\epsilon }t_{\mu _{1}\cdots \mu _{k}|\nu _{1}\cdots \nu
_{k}}=\epsilon _{\mu _{1}\cdots \mu _{k}|\left[ \nu _{2}\cdots \nu _{k},\nu
_{1}\right] }+\epsilon _{\nu _{1}\cdots \nu _{k}|\left[ \mu _{2}\cdots \mu
_{k},\mu _{1}\right] },  \label{k8}
\end{equation}
where we used the common notation $f_{,\mu }\equiv \partial _{\mu }f$.
Indeed, for $k=1$, $t_{\mu _{1}\cdots \mu _{k}|\nu _{1}\cdots \nu _{k}}$
becomes a symmetric two-tensor field, traditionally denoted by $h_{\mu \nu }$
(the Pauli-Fierz field) and (\ref{k8}) takes the familiar form $\delta
_{\epsilon }h_{\mu \nu }=\partial _{\mu }\epsilon _{\nu }+\partial _{\nu
}\epsilon _{\mu }$, while for $k=2$ one gets a tensor field $t_{\mu \nu
|\alpha \beta }$ with the mixed symmetry of the Riemann tensor, subject to
the gauge transformations $\delta _{\epsilon }t_{\mu \nu |\alpha \beta
}=\epsilon _{\mu \nu |\left[ \beta ,\alpha \right] }+\epsilon _{\alpha \beta
|\left[ \nu ,\mu \right] }$. The mixed symmetry properties of the gauge
parameters $\epsilon _{\mu _{1}\cdots \mu _{k}|\nu _{1}\cdots \nu _{k-1}}$
follow from those of the left-hand side of (\ref{k8}) once we require that
the mixed symmetry of $t_{\mu _{1}\cdots \mu _{k}|\nu _{1}\cdots \nu _{k}}$
is inherited by its gauge variation. As a consequence we find that $\epsilon
_{\mu _{1}\cdots \mu _{k}|\nu _{1}\cdots \nu _{k-1}}$ displays the mixed
symmetry $\left( k,k-1\right) $%
\begin{equation}
\epsilon _{\mu _{1}\cdots \mu _{k}|\nu _{1}\cdots \nu _{k-1}}= 
\begin{array}{ll}
\mu _{1} & \nu _{1} \\ 
\vdots & \vdots \\ 
\vdots & \nu _{k-1} \\ 
\mu _{k} & 
\end{array}
,  \label{k8a}
\end{equation}
so it is separately antisymmetric in the first $k$ and respectively last $%
\left( k-1\right) $ indices and satisfies the identity 
\begin{equation}
\epsilon _{\left[ \mu _{1}\cdots \mu _{k}|\nu _{1}\right] \nu _{2}\cdots \nu
_{k-1}}\equiv 0.  \label{k9}
\end{equation}
The formula (\ref{k9}) has the role to enforce that $\delta _{\epsilon
}t_{\mu _{1}\cdots \mu _{k}|\nu _{1}\cdots \nu _{k}}$ satisfies a Bianchi I
identity similar to that of the field itself, namely, (\ref{k3}). Taking
into account the gauge transformations (\ref{k8}), we find that the most
general form of a Lagrangian action that complies with all the above
mentioned requirements is given by 
\begin{eqnarray}
&&S^{\mathrm{L}}\left[ t_{\mu _{1}\cdots \mu _{k}|\nu _{1}\cdots \nu
_{k}}\right] =c_{1}\int d^{D}x\left[ \frac{\left( -\right) ^{k+1}}{2}\left(
\partial _{\rho }t\right) \left( \partial ^{\rho }t\right) +\right. 
\nonumber \\
&&\sum_{m=0}^{k-1}\left( -\right) ^{m}\left( ^{k}C_{m}\right) ^{2}\left( -%
\frac{1}{2}\left( \partial _{\rho }t_{\mu _{1}\cdots \mu _{k-m}|\nu _{1}\nu
_{2}\cdots \nu _{k-m}}\right) \left( \partial ^{\rho }t^{\mu _{1}\cdots \mu
_{k-m}|\nu _{1}\nu _{2}\cdots \nu _{k-m}}\right) +\right.  \nonumber \\
&&\left( k-m\right) \left( \partial ^{\rho }t_{\rho \mu _{1}\cdots \mu
_{k-m-1}|\nu _{1}\nu _{2}\cdots \nu _{k-m}}\right) \left( \partial _{\alpha
}t^{\alpha \mu _{1}\cdots \mu _{k-m-1}|\nu _{1}\nu _{2}\cdots \nu
_{k-m}}\right) -  \nonumber \\
&&\left. \left. \frac{\left( k-m\right) ^{2}}{m+1}\left( \partial ^{\rho
}t_{\rho \mu _{1}\cdots \mu _{k-m-1}|\nu _{1}\nu _{2}\cdots \nu
_{k-m}}\right) \left( \partial ^{\nu _{1}}t^{\nu _{2}\cdots \nu _{k-m}|\mu
_{1}\cdots \mu _{k-m-1}}\right) \right) \right] ,  \label{k10}
\end{eqnarray}
where $^{k}C_{m}$ represents the number of combinations of $m$ objects drawn
from $k$ and $c_{1}$ is a non-vanishing real constant. If we conveniently
fix the value of this constant to 
\begin{equation}
c_{1}=\frac{\left( -\right) ^{k+1}}{k^{2}},  \label{k11}
\end{equation}
then for $k=1,2$ we are led precisely to the Lagrangian actions \cite
{pf,zinov1} corresponding to the Pauli-Fierz field and respectively to the
tensor field with the mixed symmetry of the Riemann tensor.

It can be checked that (\ref{k8}) is a generating set of gauge
transformations for the action (\ref{k10}). This generating set is abelian
and off-shell $\left( k-1\right) $-order reducible. Indeed, if we make the
transformation 
\begin{eqnarray}
\epsilon _{\mu _{1}\cdots \mu _{k}|\nu _{1}\cdots \nu _{k-1}} &=&\partial
_{\left[ \mu _{1}\right. }\stackrel{(1)}{\chi }_{\left. \mu _{2}\cdots \mu
_{k}\right] \left[ \nu _{1}|\nu _{2}\cdots \nu _{k-1}\right] }  \nonumber \\
&&+\left( -\right) ^{k+1}2\stackrel{(1)}{\chi }_{\mu _{1}\cdots \mu
_{k}|\left[ \nu _{2}\cdots \nu _{k-1},\nu _{1}\right] },  \label{k12}
\end{eqnarray}
with $\stackrel{(1)}{\chi }_{\mu _{1}\cdots \mu _{k}|\nu _{1}\cdots \nu
_{k-2}}$ an arbitrary tensor field on $\mathcal{M}$ displaying the mixed
symmetry $\left( k,k-2\right) $, then the gauge transformations of the
tensor field vanish identically 
\begin{equation}
\delta _{\epsilon \left( \stackrel{(1)}{\chi }\right) }t_{\mu _{1}\cdots \mu
_{k}|\nu _{1}\cdots \nu _{k}}=0.  \label{k13}
\end{equation}
Next, if we perform the transformation 
\begin{eqnarray}
\stackrel{(1)}{\chi }_{\mu _{1}\cdots \mu _{k}|\nu _{1}\cdots \nu _{k-2}}
&=&\partial _{\left[ \mu _{1}\right. }\stackrel{(2)}{\chi }_{\left. \mu
_{2}\cdots \mu _{k}\right] \left[ \nu _{1}|\nu _{2}\cdots \nu _{k-2}\right] }
\nonumber \\
&&+\left( -\right) ^{k+1}3\stackrel{(2)}{\chi }_{\mu _{1}\cdots \mu
_{k}|\left[ \nu _{2}\cdots \nu _{k-2},\nu _{1}\right] },  \label{k14}
\end{eqnarray}
with $\stackrel{(2)}{\chi }_{\mu _{1}\cdots \mu _{k}|\nu _{1}\cdots \nu
_{k-3}}$ an arbitrary tensor field on $\mathcal{M}$ that exhibits the mixed
symmetry $\left( k,k-3\right) $, then we find that the gauge transformed
parameters (\ref{k12}) strongly vanish 
\begin{equation}
\epsilon _{\mu _{1}\cdots \mu _{k}|\nu _{1}\cdots \nu _{k-1}}\left( 
\stackrel{(1)}{\chi }\left( \stackrel{(2)}{\chi }\right) \right) =0.
\label{k15}
\end{equation}
Along a similar line it can be shown that if we perform the changes 
\begin{eqnarray}
\stackrel{(m)}{\chi }_{\mu _{1}\cdots \mu _{k}|\nu _{1}\cdots \nu _{k-m-1}}
&=&\partial _{\left[ \mu _{1}\right. }\stackrel{(m+1)}{\chi }_{\left. \mu
_{2}\cdots \mu _{k}\right] \left[ \nu _{1}|\nu _{2}\cdots \nu
_{k-m-1}\right] }  \nonumber \\
&&+\left( -\right) ^{k+1}\left( m+2\right) \stackrel{(m+1)}{\chi }_{\mu
_{1}\cdots \mu _{k}|\left[ \nu _{2}\cdots \nu _{k-m-1},\nu _{1}\right] },
\label{k16}
\end{eqnarray}
for $2\leq m\leq k-2$, with $\stackrel{(m)}{\chi }_{\mu _{1}\cdots \mu
_{k}|\nu _{1}\cdots \nu _{k-m-1}}$ some arbitrary tensor fields on $\mathcal{%
M}$, with the mixed symmetry $\left( k,k-m-1\right) $ and $\stackrel{(k-1)}{%
\chi }_{\mu _{1}\cdots \mu _{k}|\nu _{1}\nu _{0}}\equiv \stackrel{(k-1)}{%
\chi }_{\mu _{1}\cdots \mu _{k}}$ a completely antisymmetric tensor ($k$%
-form field), then 
\begin{equation}
\stackrel{(m-1)}{\chi }_{\mu _{1}\cdots \mu _{k}|\nu _{1}\cdots \nu
_{k-m}}\left( \stackrel{(m)}{\chi }\left( \stackrel{(m+1)}{\chi }\right)
\right) =0,\;2\leq m\leq k-2,  \label{k17}
\end{equation}
while 
\begin{equation}
\stackrel{(k-2)}{\chi }_{\mu _{1}\cdots \mu _{k}|\nu _{1}}\left( \stackrel{%
(k-1)}{\chi }\right) =0  \label{k18}
\end{equation}
if and only if 
\begin{equation}
\stackrel{(k-1)}{\chi }_{\mu _{1}\cdots \mu _{k}}=0.  \label{k19}
\end{equation}
The tensor fields $\left( \stackrel{(m)}{\chi }_{\mu _{1}\cdots \mu _{k}|\nu
_{1}\cdots \nu _{k-m-1}}\right) _{m=\overline{1,k-1}}$ will be called
reducibility parameters of order $m$. Excepting $\stackrel{(k-1)}{\chi }%
_{\mu _{1}\cdots \mu _{k}}$, which is completely antisymmetric, the
reducibility parameters $\left( \stackrel{(m)}{\chi }_{\mu _{1}\cdots \mu
_{k}|\nu _{1}\cdots \nu _{k-m-1}}\right) _{m=\overline{1,k-2}}$ transform
according to some irreducible representation of $GL\left( D,\mathbb{R}%
\right) $, associated with the two-column Young diagrams 
\begin{equation}
\stackrel{(m)}{\chi }_{\mu _{1}\cdots \mu _{k}|\nu _{1}\cdots \nu _{k-m-1}}= 
\begin{array}{ll}
\mu _{1} & \nu _{1} \\ 
\vdots & \vdots \\ 
\vdots & \nu _{k-m-1} \\ 
\vdots &  \\ 
\mu _{k} & 
\end{array}
.  \label{k20}
\end{equation}
Hence, they are separately antisymmetric in the first $k$ and respectively
last $\left( k-m-1\right) $ indices and satisfy the identities 
\begin{equation}
\stackrel{(m)}{\chi }_{\left[ \mu _{1}\cdots \mu _{k}|\nu _{1}\right] \nu
_{2}\cdots \nu _{k-m-1}}\equiv 0.  \label{k21}
\end{equation}
In view of the reducibility structure exhibited by the generating set (\ref
{k8}) of gauge transformations, it follows that the spacetime dimension is
subject to the condition 
\begin{equation}
D\geq 2k+1,  \label{dimsptime}
\end{equation}
such that the $\left( k,k\right) $ tensor field has a non-negative number of
physical degrees of freedom. [For $D=2k$ the Lagrangian action (\ref{k10})
reduces to a full divergence, for $D=2k+1$ the $\left( k,k\right) $ tensor
field has zero physical degrees of freedom, while for $D>2k+1$ it possesses
strictly positive values of the number of physical degrees of freedom.]

The field equations resulting from the action (\ref{k10}) 
\begin{equation}
\frac{\delta S^{\mathrm{L}}\left[ t_{\mu _{1}\cdots \mu _{k}|\nu _{1}\cdots
\nu _{k}}\right] }{\delta t_{\mu _{1}\cdots \mu _{k}|\nu _{1}\cdots \nu _{k}}%
}\equiv c_{1}T^{\mu _{1}\cdots \mu _{k}|\nu _{1}\cdots \nu _{k}}\approx 0,
\label{k22}
\end{equation}
involve the tensor $T^{\mu _{1}\cdots \mu _{k}|\nu _{1}\cdots \nu _{k}}$,
which is linear in the tensor field $t^{\mu _{1}\cdots \mu _{k}|\nu
_{1}\cdots \nu _{k}}$, second-order in its derivatives and displays the
mixed symmetry $\left( k,k\right) $. Its concrete expression reads as 
\begin{eqnarray}
&&T_{\;\;\;\;\;\;\;\;\;\;\nu _{1}\cdots \nu _{k}}^{\mu _{1}\cdots \mu
_{k}|}=\Box t_{\;\;\;\;\;\;\;\;\;\;\nu _{1}\cdots \nu _{k}}^{\mu _{1}\cdots
\mu _{k}|}+\left( -\right) ^{k}\left( \partial _{\rho }\partial ^{\left[ \mu
_{1}\right. }t_{\;\;\;\;\;\;\;\;\;\;\;\;\;\;\nu _{1}\cdots \nu _{k}}^{\left.
\mu _{2}\cdots \mu _{k}\right] \rho |}\right.  \nonumber \\
&&\left. +\partial ^{\rho }\partial _{\left[ \nu _{1}\right. }t_{\left. \nu
_{2}\cdots \nu _{k}\right] \rho |}^{\;\;\;\;\;\;\;\;\;\;\;\;\mu _{1}\cdots
\mu _{k}}\right) +\partial ^{\left[ \mu _{1}\right.
}t_{\;\;\;\;\;\;\;\;\;\;\;\;\left[ \nu _{2}\cdots \nu _{k},\nu _{1}\right]
}^{\left. \mu _{2}\cdots \mu _{k}\right] |}  \nonumber \\
&&+\partial _{\rho }\partial ^{\lambda }\left( \sum_{m=1}^{k}\left( \frac{%
\left( -\right) ^{m}}{m!}\delta _{\left[ \nu _{1}\right. }^{\left[ \mu
_{1}\right. }\cdots \delta _{\nu _{m}}^{\mu _{m}}\left(
t_{\;\;\;\;\;\;\;\;\;\;\;\;\left. \nu _{m+1}\cdots \nu _{k}\right] }^{\left.
\mu _{m+1}\cdots \mu _{k}\right] |}\delta _{\lambda }^{\rho
}-mt_{\;\;\;\;\;\;\;\;\;\;\;\;\left. \nu _{m+1}\cdots \nu _{k}\right]
\lambda }^{\left. \mu _{m+1}\cdots \mu _{k}\right] \rho |}\right) \right)
\right)  \nonumber \\
&&+\sum_{m=1}^{k-1}\left( \frac{1}{m!}\delta _{\left[ \nu _{1}\right.
}^{\left[ \mu _{1}\right. }\cdots \delta _{\nu _{m}}^{\mu _{m}}\left( \left(
-\right) ^{k}\left( \partial _{{}}^{\mu
_{m+1}}t_{\;\;\;\;\;\;\;\;\;\;\;\;\;\;\left. \nu _{m+1}\cdots \nu
_{k}\right] ,\rho }^{\left. \mu _{m+2}\cdots \mu _{k}\right] \rho |}\right.
\right. \right.  \nonumber \\
&&\left. \left. \left. +\partial _{\nu _{m+1}}^{{}}t_{\left. \nu
_{m+2}\cdots \nu _{k}\right] \rho |}^{\;\;\;\;\;\;\;\;\;\;\;\;\;\;\left. \mu
_{m+1}\cdots \mu _{k}\right] ,\rho }\right) +\frac{\left( -\right) ^{m}}{m+1}%
\partial ^{\mu _{m+1}}t_{\;\;\;\;\;\;\;\;\;\;\;\;\;\left. \nu _{m+2}\cdots
\nu _{k},\nu _{m+1}\right] }^{\left. \mu _{m+2}\cdots \mu _{k}\right]
|}\right) \right) .  \label{k23}
\end{eqnarray}
We notice that our antisymmetrization convention takes into account only the
distinct terms. For instance, in 
\[
\delta _{\left[ \nu _{1}\right. }^{\left[ \mu _{1}\right. }\cdots \delta
_{\nu _{m}}^{\mu _{m}}t_{\;\;\;\;\;\;\;\;\;\;\;\;\left. \nu _{m+1}\cdots \nu
_{k}\right] }^{\left. \mu _{m+1}\cdots \mu _{k}\right] |} 
\]
it is understood that there appear only $\left( k!/\left( k-m\right)
!\right) ^{2}/m!$ terms. In a somehow abusive language we will name the
components of this tensor the Euler-Lagrange (E.L.) derivatives of the
action (\ref{k10}). The various traces of $T^{\mu _{1}\cdots \mu _{k}|\nu
_{1}\cdots \nu _{k}}$ will be denoted by $\left( T^{\mu _{m+1}\cdots \mu
_{k}|\nu _{m+1}\cdots \nu _{k}}\right) _{m=\overline{1,k}}$, being
understood that they are defined in a manner similar to (\ref{k5}). The
gauge invariance of the Lagrangian action (\ref{k10}) under the
transformations (\ref{k8}) is equivalent to the fact that the functions
defining the field equations are not all independent, but rather obey the
Noether identities 
\begin{equation}
\partial _{\mu }\frac{\delta S^{\mathrm{L}}\left[ t_{\mu _{1}\cdots \mu
_{k}|\nu _{1}\cdots \nu _{k}}\right] }{\delta t_{\mu \mu _{1}\cdots \mu
_{k-1}|\nu _{1}\cdots \nu _{k}}}\equiv c_{1}\partial _{\mu }T^{\mu \mu
_{1}\cdots \mu _{k-1}|\nu _{1}\cdots \nu _{k}}=0,  \label{k24}
\end{equation}
while the presence of the reducibility shows that not all of the above
Noether identities are independent. It can be checked that the functions (%
\ref{k23}) defining the field equations, the gauge generators, as well as
all the reducibility functions, satisfy the general regularity assumptions
from~\cite{genreg}, such that the model under discussion is described by a
normal gauge theory of Cauchy order equal to $\left( k+1\right) $.

\section{Reconstruction of the Lagrangian formulation from the generalized
3-complex\label{3}}

\subsection{Gauge invariance\label{3.1}}

This model describes a free gauge theory that can be interpreted in a
consistent manner in terms of the generalized differential complex~\cite
{genpoinc} $\Omega _{2}\left( \mathcal{M}\right) $ of tensor fields with
mixed symmetries corresponding to a maximal sequence of Young diagrams with
two columns, defined on a pseudo-Riemannian manifold $\mathcal{M}$ of
dimension $D$. Let us denote by $\bar{d}$ the associated operator
(3-differential) that is third-order nilpotent, $\bar{d}^{3}=0$, and by $%
\Omega _{2}^{p}\left( \mathcal{M}\right) $ the vector space spanned by the
tensor fields from $\Omega _{2}\left( \mathcal{M}\right) $ with $p$ entries.
The action of $\bar{d}$ on an element pertaining to $\Omega _{2}^{p}\left( 
\mathcal{M}\right) $ results in a tensor from $\Omega _{2}^{p+1}\left( 
\mathcal{M}\right) $ with one spacetime derivative, the action of $\bar{d}%
^{2}$ on a similar element leads to a tensor from $\Omega _{2}^{p+2}\left( 
\mathcal{M}\right) $ containing two spacetime derivatives, while the action
of $\bar{d}^{3}$ on any such element identically vanishes. In brief, the
generalized 3-complex $\Omega _{2}\left( \mathcal{M}\right) $ may
suggestively be represented through the commutative diagram 
\[
\begin{array}{ccccccccccc}
&  &  &  &  &  &  &  &  &  & \vdots \\ 
&  &  &  &  &  &  &  &  & \stackrel{\bar{d}^{2}}{\nearrow } & \uparrow ^{%
\bar{d}} \\ 
&  &  &  &  &  &  &  & \mathbf{\Omega }_{2}^{2k+2} & \stackrel{\bar{d}}{%
\rightarrow } & \mathbf{\Omega }_{2}^{2k+3} \\ 
&  &  &  &  &  &  & \stackrel{\bar{d}^{2}}{\nearrow } & \uparrow ^{\bar{d}}
&  &  \\ 
&  &  &  &  &  & \mathbf{\Omega }_{2}^{2k} & \stackrel{\bar{d}}{\rightarrow }
& \mathbf{\Omega }_{2}^{2k+1} &  &  \\ 
&  &  &  &  & \stackrel{\bar{d}^{2}}{\nearrow } & \uparrow ^{\bar{d}} &  & 
&  &  \\ 
&  &  &  & \vdots & \stackrel{\bar{d}}{\rightarrow } & \mathbf{\Omega }%
_{2}^{2k-1} &  &  &  &  \\ 
&  &  & \stackrel{\bar{d}^{2}}{\nearrow } & \uparrow ^{\bar{d}} &  &  &  & 
&  &  \\ 
&  & \Omega _{2}^{2} & \stackrel{\bar{d}}{\rightarrow } & \cdots &  &  &  & 
&  &  \\ 
& \stackrel{\bar{d}^{2}}{\nearrow } & \uparrow ^{\bar{d}} &  &  &  &  &  & 
&  &  \\ 
\Omega _{2}^{0} & \stackrel{\bar{d}}{\rightarrow } & \Omega _{2}^{1} &  &  & 
&  &  &  &  & 
\end{array}
\]
where the third-order nilpotency of $\bar{d}$ means that any vertical arrow
followed by the closest higher diagonal arrow maps to zero, and the same
with respect to any diagonal arrow followed by the closest higher horizontal
one. Its bold part emphasizes the sequences that apply to the model under
discussion: the first one governs the dynamics and indicates the presence of
some gauge symmetry 
\begin{equation}
\begin{array}{ccccc}
\Omega _{2}^{2k} & \stackrel{\bar{d}^{2}}{\rightarrow } & \Omega _{2}^{2k+2}
& \stackrel{\bar{d}}{\rightarrow } & \Omega _{2}^{2k+3} \\ 
\begin{array}{l}
\mathrm{field} \\ 
t_{\mu _{1}\cdots \mu _{k}|\nu _{1}\cdots \nu _{k}}
\end{array}
&  & 
\begin{array}{l}
\mathrm{curvature} \\ 
F_{\mu _{1}\cdots \mu _{k+1}|\nu _{1}\cdots \nu _{k+1}}
\end{array}
&  & 
\begin{array}{l}
\mathrm{Bianchi\;II} \\ 
\partial _{\left[ \mu _{1}\right. }F_{\left. \mu _{2}\cdots \mu
_{k+1}\right] |\nu _{1}\cdots \nu _{k+1}}\equiv 0
\end{array}
\end{array}
,  \label{k25}
\end{equation}
while the second sequence solves the gauge symmetry 
\begin{equation}
\begin{array}{ccccc}
\Omega _{2}^{2k-1} & \stackrel{\bar{d}}{\rightarrow } & \Omega _{2}^{2k} & 
\stackrel{\bar{d}^{2}}{\rightarrow } & \Omega _{2}^{2k+2} \\ 
\begin{array}{l}
\mathrm{gauge\;param.} \\ 
\epsilon _{\mu _{1}\cdots \mu _{k}|\nu _{1}\cdots \nu _{k-1}}
\end{array}
&  & 
\begin{array}{l}
\mathrm{gauge\;transf.} \\ 
\delta _{\epsilon }t_{\mu _{1}\cdots \mu _{k}|\nu _{1}\cdots \nu _{k}}
\end{array}
&  & 
\begin{array}{l}
\mathrm{gauge\;inv.\;objects} \\ 
\delta _{\epsilon }F_{\mu _{1}\cdots \mu _{k+1}|\nu _{1}\cdots \nu _{k+1}}=0
\end{array}
\end{array}
.  \label{k26}
\end{equation}
Let us discuss the previous sequences. Starting from the tensor field $%
t_{\mu _{1}\cdots \mu _{k}|\nu _{1}\cdots \nu _{k}}$ from $\Omega _{2}^{2k}$%
, we can construct its curvature tensor $F_{\mu _{1}\cdots \mu _{k+1}|\nu
_{1}\cdots \nu _{k+1}}$, defined via 
\begin{equation}
\left( \bar{d}^{2}t\right) _{\mu _{1}\cdots \mu _{k+1}\nu _{1}\cdots \nu
_{k+1}}\sim F_{\mu _{1}\cdots \mu _{k+1}|\nu _{1}\cdots \nu _{k+1}}=\partial
_{\left[ \mu _{1}\right. }t_{\left. \mu _{2}\cdots \mu _{k+1}\right] |\left[
\nu _{2}\cdots \nu _{k+1},\nu _{1}\right] },  \label{k27}
\end{equation}
which is second-order in the spacetime derivatives and belongs to $\Omega
_{2}^{2k+2}$. Thus, the curvature tensor transforms in an irreducible
representation of $GL\left( D,\mathbb{R}\right) $ and exhibits the
symmetries of the rectangular two-column Young diagram 
\begin{equation}
F_{\mu _{1}\cdots \mu _{k+1}|\nu _{1}\cdots \nu _{k+1}}= 
\begin{array}{cc}
\mu _{1} & \nu _{1} \\ 
\vdots & \vdots \\ 
\mu _{k+1} & \nu _{k+1}
\end{array}
,  \label{k28}
\end{equation}
being separately antisymmetric in the first and respectively last $\left(
k+1\right) $ indices, symmetric under the inter-change between the two sets
of indices 
\begin{equation}
F_{\mu _{1}\cdots \mu _{k+1}|\nu _{1}\cdots \nu _{k+1}}=F_{\nu _{1}\cdots
\nu _{k+1}|\mu _{1}\cdots \mu _{k+1}}  \label{k29}
\end{equation}
and obeying the (algebraic) Bianchi I identity 
\begin{equation}
F_{\left[ \mu _{1}\cdots \mu _{k+1}|\nu _{1}\right] \nu _{2}\cdots \nu
_{k+1}}\equiv 0.  \label{k30}
\end{equation}
The action of $\bar{d}$ on $F_{\mu _{1}\cdots \mu _{k+1}|\nu _{1}\cdots \nu
_{k+1}}$ maps to zero 
\begin{eqnarray}
\left( \bar{d}^{3}t\right) _{\mu _{1}\cdots \mu _{k+2}\nu _{1}\cdots \nu
_{k+1}} &=&\left( \bar{d}F\right) _{\mu _{1}\cdots \mu _{k+2}\nu _{1}\cdots
\nu _{k+1}}  \nonumber \\
&\sim &\partial _{\left[ \mu _{1}\right. }F_{\left. \mu _{2}\cdots \mu
_{k+2}\right] |\nu _{1}\cdots \nu _{k+1}}\equiv 0,  \label{k31}
\end{eqnarray}
and represents nothing but the (differential) Bianchi II identity for the
curvature tensor. Since the curvature tensor and its traces \textit{are the
most general non-vanishing second-order derivative quantities in }$\Omega
_{2}\left( \mathcal{M}\right) $\textit{\ constructed from }$t_{\mu
_{1}\cdots \mu _{k}|\nu _{1}\cdots \nu _{k}}$\textit{, }we expect that the
E.L. derivatives of the action, (\ref{k23}), completely rely on it. The
formula (\ref{k24}) shows that the corresponding field equations cannot be
all independent, but satisfy some Noether identities related to the Bianchi
II identity of the curvature tensor. This already points out that the
searched for free Lagrangian action must be invariant under a certain gauge
symmetry. The second sequence, namely (\ref{k26}), gives the form of the
gauge invariance. As the free field equations involve $F_{\mu _{1}\cdots \mu
_{k+1}|\nu _{1}\cdots \nu _{k+1}}$, it is natural to require that these are
the most general gauge invariant quantities 
\begin{equation}
\delta _{\epsilon }\left( \bar{d}^{2}t\right) _{\mu _{1}\cdots \mu _{k+1}\nu
_{1}\cdots \nu _{k+1}}\sim \delta _{\epsilon }F_{\mu _{1}\cdots \mu
_{k+1}|\nu _{1}\cdots \nu _{k+1}}=0.  \label{k32}
\end{equation}
This matter is immediately solved if we take 
\begin{eqnarray}
\left( \bar{d}\epsilon \right) _{\mu _{1}\cdots \mu _{k}|\nu _{1}\cdots \nu
_{k}} &\sim &\epsilon _{\mu _{1}\cdots \mu _{k}|\left[ \nu _{2}\cdots \nu
_{k},\nu _{1}\right] }+\epsilon _{\nu _{1}\cdots \nu _{k}|\left[ \mu
_{2}\cdots \mu _{k},\mu _{1}\right] }  \nonumber \\
&=&\delta _{\epsilon }t_{\mu _{1}\cdots \mu _{k}|\nu _{1}\cdots \nu _{k}},
\label{k33}
\end{eqnarray}
where the gauge parameters $\epsilon _{\mu _{1}\cdots \mu _{k}|\nu
_{1}\cdots \nu _{k-1}}$ pertain to $\Omega _{2}^{2k-1}$, because, on account
of the third-order nilpotency of $\bar{d}$, we find that 
\begin{equation}
\delta _{\epsilon }F_{\mu _{1}\cdots \mu _{k+1}|\nu _{1}\cdots \nu
_{k+1}}\sim \left( \bar{d}^{3}\epsilon \right) _{\mu _{1}\cdots \mu
_{k+1}\nu _{1}\cdots \nu _{k+1}}\equiv 0.  \label{k34}
\end{equation}
Clearly, the relation (\ref{k33}) coincides with the gauge transformations (%
\ref{k8}).

\subsection{Lagrangian action\label{3.2}}

We complete our discussion by exemplifying the construction of the free
field equations. Let us denote by $S^{\prime \mathrm{L}}\left[ t_{\mu
_{1}\cdots \mu _{k}|\nu _{1}\cdots \nu _{k}}\right] $ a free, second-order
derivative action that is gauge invariant under (\ref{k8}), and by $\delta
S^{\prime \mathrm{L}}/\delta t^{\mu _{1}\cdots \mu _{k}|\nu _{1}\cdots \nu
_{k}}$ its functional derivatives with respect to the fields, which are
imposed to depend linearly on the undifferentiated curvature tensor. Then,
as these functional derivatives must have the same mixed symmetry like $%
t_{\mu _{1}\cdots \mu _{k}|\nu _{1}\cdots \nu _{k}}$, it follows that they
necessarily determine a tensor from $\Omega _{2}^{2k}$. The operations that
can be performed with respect to the curvature tensor in order to reduce its
number of indices without increasing its derivative order is to take its
(multiple) traces 
\begin{equation}
F_{\mu _{m+2}\cdots \mu _{k+1}|\nu _{m+2}\cdots \nu _{k+1}}=\sigma ^{\mu
_{1}\nu _{1}}\cdots \sigma ^{\mu _{m+1}\nu _{m+1}}F_{\mu _{1}\cdots \mu
_{k+1}|\nu _{1}\cdots \nu _{k+1}}\in \Omega _{2}^{2\left( k-m\right) },\;m=%
\overline{0,k},  \label{k35}
\end{equation}
being understood that we maintain the conventions (\ref{conv}). By direct
computation, from (\ref{k27}) we get that the various traces of the
curvature tensor have the expressions 
\begin{eqnarray}
&&F_{\mu _{1}\cdots \mu _{k}|\nu _{1}\cdots \nu _{k}}=\Box t_{\mu _{1}\cdots
\mu _{k}|\nu _{1}\cdots \nu _{k}}+\left( -\right) ^{k}\partial ^{\rho
}\left( \partial _{\left[ \mu _{1}\right. }t_{\left. \mu _{2}\cdots \mu
_{k}\right] \rho |\nu _{1}\cdots \nu _{k}}\right.  \nonumber \\
&&\left. +\partial _{\left[ \nu _{1}\right. }t_{\left. \nu _{2}\cdots \nu
_{k}\right] \rho |\mu _{1}\cdots \mu _{k}}\right) +\partial _{\left[ \mu
_{1}\right. }t_{\left. \mu _{2}\cdots \mu _{k}\right] |\left[ \nu _{2}\cdots
\nu _{k},\nu _{1}\right] },  \label{k36}
\end{eqnarray}
\begin{eqnarray}
&&F_{\mu _{1}\cdots \mu _{k-m}|\nu _{1}\cdots \nu _{k-m}}=\left( m+1\right)
\Box t_{\mu _{1}\cdots \mu _{k-m}|\nu _{1}\cdots \nu _{k-m}}+\partial
_{\left[ \mu _{1}\right. }t_{\left. \mu _{2}\cdots \mu _{k-m}\right] |\left[
\nu _{2}\cdots \nu _{k-m},\nu _{1}\right] }  \nonumber \\
&&+\left( -\right) ^{k+m}\left( m+1\right) \partial ^{\rho }\left( \partial
_{\left[ \mu _{1}\right. }t_{\left. \mu _{2}\cdots \mu _{k-m}\right] \rho
|\nu _{1}\cdots \nu _{k-m}}+\partial _{\left[ \nu _{1}\right. }t_{\left. \nu
_{2}\cdots \nu _{k-m}\right] \rho |\mu _{1}\cdots \mu _{k-m}}\right) 
\nonumber \\
&&-m\left( m+1\right) \partial ^{\rho }\partial ^{\lambda }t_{\mu _{1}\cdots
\mu _{k-m}\rho |\nu _{1}\cdots \nu _{k-m}\lambda },\;m=\overline{1,k-1},
\label{k37}
\end{eqnarray}
\begin{equation}
F=\left( k+1\right) \Box t-k\left( k+1\right) \partial ^{\rho }\partial
^{\lambda }t_{\rho |\lambda },  \label{k38}
\end{equation}
where $t_{\rho |\lambda }$ is a symmetric two-tensor. The only (linear)
combinations formed with these quantities that belong to $\Omega _{2}^{2k}$
are generated by $F_{\mu _{1}\cdots \mu _{k}|\nu _{1}\cdots \nu _{k}}$ and $%
\left( \stackrel{(m)}{M}_{\mu _{1}\cdots \mu _{k}|\nu _{1}\cdots \nu
_{k}}\right) _{m=\overline{1,k}}$, where 
\begin{equation}
\stackrel{(m)}{M}_{\;\;\;\;\;\;\;\;\;\;\nu _{1}\cdots \nu _{k}}^{\mu
_{1}\cdots \mu _{k}|}\equiv \delta _{\left[ \nu _{1}\right. }^{\left[ \mu
_{1}\right. }\cdots \delta _{\nu _{m}}^{\mu
_{m}}F_{\;\;\;\;\;\;\;\;\;\;\left. \nu _{m+1}\cdots \nu _{k}\right]
}^{\left. \mu _{m+1}\cdots \mu _{k}\right] |},\;m=\overline{1,k},
\label{k39}
\end{equation}
so in principle $\delta S^{\mathrm{L}}/\delta t^{\mu _{1}\cdots \mu _{k}|\nu
_{1}\cdots \nu _{k}}$ can be written as a linear combination of these
objects with coefficients that are real constants 
\begin{equation}
\frac{\delta S^{\prime \mathrm{L}}}{\delta t^{\mu _{1}\cdots \mu _{k}|\nu
_{1}\cdots \nu _{k}}}=c_{1}F_{\mu _{1}\cdots \mu _{k}|\nu _{1}\cdots \nu
_{k}}+\sum_{m=1}^{k}c_{m+1}\stackrel{(m)}{M}_{\mu _{1}\cdots \mu _{k}|\nu
_{1}\cdots \nu _{k}}.  \label{k40}
\end{equation}
According to our antisymmetrization convention, the right-hand side of (\ref
{k39}) contains only the independent terms, in number of $\left( k!/\left(
k-m\right) !\right) ^{2}/m!$. However, the requirement that the above linear
combination indeed stands for the functional derivatives of a sole
functional restricts the parametrization of the functional derivatives by
means of one constant only 
\begin{equation}
c_{m+1}=\frac{\left( -\right) ^{m}}{\left( m+1\right) !}c_{1},\;m=\overline{%
1,k},  \label{k41}
\end{equation}
so we finally find that 
\begin{equation}
\frac{\delta S^{\prime \mathrm{L}}}{\delta t^{\mu _{1}\cdots \mu _{k}|\nu
_{1}\cdots \nu _{k}}}=c_{1}\left( F_{\mu _{1}\cdots \mu _{k}|\nu _{1}\cdots
\nu _{k}}+\sum_{m=1}^{k}\frac{\left( -\right) ^{m}}{\left( m+1\right) !}%
\stackrel{(m)}{M}_{\mu _{1}\cdots \mu _{k}|\nu _{1}\cdots \nu _{k}}\right) ,
\label{k42}
\end{equation}
where $\stackrel{(m)}{M}_{\mu _{1}\cdots \mu _{k}|\nu _{1}\cdots \nu _{k}}$
are defined in (\ref{k39}). By taking into account the formulas (\ref{k36}--%
\ref{k38}) we observe that from (\ref{k42}) we precisely recover the
Lagrangian action (\ref{k10}) together with the field equations (\ref{k22}).
The E.L. derivatives (\ref{k42}) coincide with (\ref{k23}) up to the
numerical factor $c_{1}$. This also allows us to identify the expression of $%
T_{\mu _{1}\cdots \mu _{k}|\nu _{1}\cdots \nu _{k}}$ from (\ref{k22}--\ref
{k23}) in terms of the curvature tensor like 
\begin{equation}
T_{\mu _{1}\cdots \mu _{k}|\nu _{1}\cdots \nu _{k}}=F_{\mu _{1}\cdots \mu
_{k}|\nu _{1}\cdots \nu _{k}}+\sum_{m=1}^{k}\frac{\left( -\right) ^{m}}{%
\left( m+1\right) !}\stackrel{(m)}{M}_{\mu _{1}\cdots \mu _{k}|\nu
_{1}\cdots \nu _{k}}.  \label{k43}
\end{equation}

\subsection{Relationship with the curvature tensor\label{3.3}}

At this point, we can easily see the relationship of the field equations (%
\ref{k22}) and their Noether identities (\ref{k24}) with the curvature
tensor (\ref{k27}) and accompanying Bianchi II identity (\ref{k31}). First,
we observe that the field equations (\ref{k22}) are completely equivalent
with the vanishing of the simple trace of the curvature tensor 
\begin{equation}
T_{\mu _{1}\cdots \mu _{k}|\nu _{1}\cdots \nu _{k}}\approx
0\Longleftrightarrow F_{\mu _{1}\cdots \mu _{k}|\nu _{1}\cdots \nu
_{k}}\approx 0.  \label{k44}
\end{equation}
The direct statement holds due to the fact that $T_{\mu _{1}\cdots \mu
_{k}|\nu _{1}\cdots \nu _{k}}$ is expressed only through $F_{\mu _{1}\cdots
\mu _{k}|\nu _{1}\cdots \nu _{k}}$ and its traces, such that its vanishing
implies $F_{\mu _{1}\cdots \mu _{k}|\nu _{1}\cdots \nu _{k}}\approx 0$. The
converse implication holds because the vanishing of the components $%
\stackrel{(m)}{M}_{\mu _{1}\cdots \mu _{k}|\nu _{1}\cdots \nu _{k}}$ in the
right-hand side of (\ref{k43}) is a simple consequence of $F_{\mu _{1}\cdots
\mu _{k}|\nu _{1}\cdots \nu _{k}}\approx 0$. Second, the Noether identities (%
\ref{k24}) are a direct consequence of the Bianchi II identity for the
curvature tensor 
\begin{equation}
\partial _{\left[ \rho _{1}\right. }F_{\left. \rho _{2}\nu _{1}\cdots \nu
_{k}\right] |\lambda _{1}\lambda _{2}\mu _{1}\cdots \mu _{k-1}}\equiv
0\Rightarrow \partial ^{\mu }T_{\mu \mu _{1}\cdots \mu _{k-1}|\nu _{1}\cdots
\nu _{k}}\equiv 0.  \label{k45}
\end{equation}
Indeed, on the one hand the relation (\ref{k43}) yields 
\begin{eqnarray}
&&\partial _{\mu }T_{\;\;\;\;\;\;\;\;\;\;\;\;\nu _{1}\cdots \nu _{k}}^{\mu
\mu _{1}\cdots \mu _{k-1}|}=\partial _{\mu }F_{\;\;\;\;\;\;\;\;\;\;\;\;\nu
_{1}\cdots \nu _{k}}^{\mu \mu _{1}\cdots \mu _{k-1}|}-\frac{1}{2}\partial
_{\left[ \nu _{1}\right. }F_{\left. \nu _{2}\cdots \nu _{k}\right]
|}^{\;\;\;\;\;\;\;\;\;\;\;\;\mu _{1}\cdots \mu _{k-1}}  \nonumber \\
&&+\sum_{m=2}^{k}\left( \frac{1}{m!}\delta _{\left[ \nu _{1}\right.
}^{\left[ \mu _{1}\right. }\cdots \delta _{\nu _{m-1}}^{\mu _{m-1}}\left(
\left( -\right) ^{k+m}F_{\;\;\;\;\;\;\;\;\;\;\;\;\;\;\left. \nu _{m}\cdots
\nu _{k}\right] ,\mu }^{\left. \mu _{m}\cdots \mu _{k-1}\right] \mu
|}\right. \right.  \nonumber \\
&&\left. \left. -\frac{1}{m+1}\partial _{\nu _{m}}F_{\left. \nu _{m+1}\cdots
\nu _{k}\right] |}^{\;\;\;\;\;\;\;\;\;\;\;\;\left. \mu _{m}\cdots \mu
_{k-1}\right] }\right) \right) .  \label{k46}
\end{eqnarray}
On the other hand, straightforward computation leads to 
\begin{eqnarray}
&&\sigma ^{\rho _{1}\lambda _{1}}\sigma ^{\rho _{2}\lambda _{2}}\partial
_{\left[ \rho _{1}\right. }F_{\left. \rho _{2}\nu _{1}\cdots \nu _{k}\right]
|\lambda _{1}\lambda _{2}\mu _{1}\cdots \mu _{k-1}}=  \nonumber \\
&&-2\left( \partial ^{\mu }F_{\mu \mu _{1}\cdots \mu _{k-1}|\nu _{1}\cdots
\nu _{k}}-\frac{1}{2}\partial _{\left[ \nu _{1}\right. }F_{\left. \nu
_{2}\cdots \nu _{k}\right] |\mu _{1}\cdots \mu _{k-1}}\right) ,  \label{k47}
\end{eqnarray}
\begin{eqnarray}
&&\sigma ^{\rho _{1}\lambda _{1}}\sigma ^{\rho _{2}\lambda _{2}}\sigma ^{\nu
_{1}\mu _{1}}\cdots \sigma ^{\nu _{m-1}\mu _{m-1}}\partial _{\left[ \rho
_{1}\right. }F_{\left. \rho _{2}\nu _{1}\cdots \nu _{k}\right] |\lambda
_{1}\lambda _{2}\mu _{1}\cdots \mu _{k-1}}=  \nonumber \\
&&\left( -\right) ^{m}\left( m+1\right) \left( \partial ^{\mu }F_{\mu \mu
_{m}\cdots \mu _{k-1}|\nu _{m}\cdots \nu _{k}}-\frac{1}{m+1}\partial
_{\left[ \nu _{m}\right. }F_{\left. \nu _{m+1}\cdots \nu _{k}\right] |\mu
_{m}\cdots \mu _{k-1}}\right) ,  \label{k48}
\end{eqnarray}
for all $m=\overline{2,k}$. Thus, according to (\ref{k47}--\ref{k48}), we
can state that the Bianchi II identity for the curvature tensor implies the
identically vanishing of the right-hand side of (\ref{k46}), and hence
enforces the Noether identities (\ref{k24}) for the action (\ref{k10}).

\subsection{Generalized cohomology of the 3-complex\label{3.4}}

Next, we point out the relation between the generalized cohomology of the
3-complex $\Omega _{2}\left( \mathcal{M}\right) $ and our model. The
generalized cohomology of the 3-complex $\Omega _{2}\left( \mathcal{M}%
\right) $ is given by the family of graded vector spaces $H_{m}\left( \bar{d}%
\right) =Ker\left( \bar{d}^{m}\right) /Im\left( \bar{d}^{3-m}\right) $, with 
$m=1,2$. Each vector space $H_{m}\left( \bar{d}\right) $ splits into the
cohomology spaces $H_{\left( m\right) }^{p}\left( \Omega _{2}\left( \mathcal{%
M}\right) \right) $, defined like the equivalence classes of tensors from $%
\Omega _{2}^{p}\left( \mathcal{M}\right) $ that are $\bar{d}^{m}$-closed,
with any two such tensors that differ by a $\bar{d}^{3-m}$-exact element in
the same equivalence class. The spaces $H_{\left( m\right) }^{p}$ are not
empty in general, even if $\mathcal{M}$ has a trivial topology. However, in
the case where $\mathcal{M}$ (assumed to be of dimension $D$) has the
topology of $\mathbb{R}^{D}$, the generalized Poincar\'{e} lemma~\cite
{genpoinc} applied to our situation states that the generalized cohomology
of the 3-differential $\bar{d}$ on tensors represented by rectangular
diagrams with two columns is empty in the space $\Omega _{2}\left( \mathbb{R}%
^{D}\right) $ of maximal two-column tensors, $H_{\left( m\right)
}^{2n}\left( \Omega _{2}\left( \mathbb{R}^{D}\right) \right) =0$, for $1\leq
n\leq D-1$ and $m=1,2$. In particular, for $n=k+1$ and $m=1$ we find that $%
H_{\left( 1\right) }^{2k+2}\left( \Omega _{2}\left( \mathbb{R}^{D}\right)
\right) =0$ and thus, if the tensor $F_{\mu _{1}\cdots \mu _{k+1}|\nu
_{1}\cdots \nu _{k+1}}$ with the mixed symmetry (\ref{k28}) of the curvature
tensor is $\bar{d}$-closed, then it is also $\bar{d}^{2}$-exact. To put it
otherwise, if this tensor satisfies the Bianchi II identity $\partial
_{\left[ \mu _{1}\right. }F_{\left. \mu _{2}\cdots \mu _{k+2}\right] |\nu
_{1}\cdots \nu _{k+1}}\equiv 0$, then there exists an element $t_{\mu
_{1}\cdots \mu _{k}|\nu _{1}\cdots \nu _{k}}$ with the mixed symmetry (\ref
{k1}), with the help of which $F_{\mu _{1}\cdots \mu _{k+1}|\nu _{1}\cdots
\nu _{k+1}}$ can precisely be written like in (\ref{k27}).

Finally, we observe that the formula (\ref{k43}) relates the functions
defining the free field equations (\ref{k22}) to the curvature tensor by a
generalized Hodge-duality. The generalized cohomology of $\bar{d}$ on $%
\Omega _{2}\left( \mathcal{M}\right) $ when $\mathcal{M}$ has the trivial
topology of $\mathbb{R}^{D}$ together with this type of generalized
Hodge-duality reveal many important features of the free model under study.
For example, if $\bar{T}_{\mu _{1}\cdots \mu _{k}|\nu _{1}\cdots \nu _{k}}$
is a covariant tensor field with the mixed symmetry of the rectangular
two-column Young diagram (\ref{k1}) and satisfies the equation 
\begin{equation}
\partial ^{\mu }\bar{T}_{\mu \mu _{1}\cdots \mu _{k-1}|\nu _{1}\cdots \nu
_{k}}=0,  \label{k49}
\end{equation}
then there exists a tensor $\bar{\Phi}_{\mu _{1}\cdots \mu _{k+1}|\nu
_{1}\cdots \nu _{k+1}}\in \Omega _{2}\left( \mathbb{R}^{D}\right) $ with the
mixed symmetry of the curvature (of the rectangular Young diagram in (\ref
{k28})), in terms of which 
\begin{equation}
\bar{T}_{\mu _{1}\cdots \mu _{k}|\nu _{1}\cdots \nu _{k}}=\partial ^{\mu
_{k+1}}\partial ^{\nu _{k+1}}\bar{\Phi}_{\mu _{1}\cdots \mu _{k+1}|\nu
_{1}\cdots \nu _{k+1}}+c\sigma _{\mu _{1}\cdots \mu _{k}|\nu _{1}\cdots \nu
_{k}},  \label{k50}
\end{equation}
with $c$ an arbitrary real constant and $\sigma _{\mu _{1}\cdots \mu
_{k}|\nu _{1}\cdots \nu _{k}}$ being defined by the complete
antisymmetrization of the product $\sigma _{\mu _{1}\nu _{1}}\cdots \sigma
_{\mu _{k}\nu _{k}}$ over the indices $\left\{ \mu _{1},\cdots ,\mu
_{k}\right\} $, which contains, according to our antisymmetrization
convention, precisely $k!$ terms. It is easy to check the above statement in
connection with the functions (\ref{k23}) that define the field equations
for the model under consideration. Indeed, direct computation provides $c=0$
and 
\begin{equation}
T_{\mu _{1}\cdots \mu _{k}|\nu _{1}\cdots \nu _{k}}=\partial ^{\mu
_{k+1}}\partial ^{\nu _{k+1}}\Phi _{\mu _{1}\cdots \mu _{k+1}|\nu _{1}\cdots
\nu _{k+1}},  \label{k51}
\end{equation}
where 
\begin{equation}
\Phi _{\;\;\;\;\;\;\;\;\;\;\;\;\nu _{1}\cdots \nu _{k+1}}^{\mu _{1}\cdots
\mu _{k+1}|}=\sum_{m=0}^{k}\frac{\left( -\right) ^{m}}{m!}\delta _{\left[
\nu _{1}\right. }^{\left[ \mu _{1}\right. }\cdots \delta _{\nu _{m}}^{\mu
_{m}}t_{\;\;\;\;\;\;\;\;\;\;\;\;\nu _{m+1}\cdots \nu _{k}}^{\mu _{m+1}\cdots
\mu _{k}|}\delta _{\left. \nu _{k+1}\right] }^{\left. \mu _{k+1}\right] },
\label{k52}
\end{equation}
such that the corresponding $\Phi _{\mu _{1}\cdots \mu _{k+1}|\nu _{1}\cdots
\nu _{k+1}}$ indeed displays the mixed symmetry (\ref{k28}) of the curvature
tensor. We note that we employed the conventions (\ref{conv}), such that the
element corresponding to $m=0$ in the right-hand side of (\ref{k52}) is 
\begin{equation}
t_{\;\;\;\;\;\;\;\;\;\;\;\;\left[ \nu _{1}\cdots \nu _{k}\right. }^{\left[
\mu _{1}\cdots \mu _{k}|\right. }\delta _{\left. \nu _{k+1}\right] }^{\left.
\mu _{k+1}\right] }  \label{k52a}
\end{equation}
and it contains precisely $\left( k+1\right) ^{2}$ terms, while the element
associated with $m=k$ reads 
\begin{equation}
\frac{\left( -\right) ^{k}}{k!}\delta _{\left[ \nu _{1}\right. }^{\left[ \mu
_{1}\right. }\cdots \delta _{\left. \nu _{k+1}\right] }^{\left. \mu
_{k+1}\right] }t,  \label{k52c}
\end{equation}
and it involves only $\left( k+1\right) !$ terms. In general, the number of
terms in 
\begin{equation}
\delta _{\left[ \nu _{1}\right. }^{\left[ \mu _{1}\right. }\cdots \delta
_{\nu _{m}}^{\mu _{m}}t_{\;\;\;\;\;\;\;\;\;\;\;\;\nu _{m+1}\cdots \nu
_{k}}^{\mu _{m+1}\cdots \mu _{k}|}\delta _{\left. \nu _{k+1}\right]
}^{\left. \mu _{k+1}\right] }  \label{k52b}
\end{equation}
for a given value of $m$ is equal to $\left( \left( k+1\right) !/\left(
k-m\right) !\right) ^{2}/\left( m+1\right) !$.

\section{BRST symmetry\label{4}}

\subsection{Construction of the differential BRST complex\label{4.1}}

In agreement with the general setting of the antibracket-antifield
formalism, the construction of the BRST symmetry for the free theory under
consideration starts with the identification of the BRST algebra on which
the BRST differential $s$ acts. The generators of the BRST algebra are of
two kinds: fields/ghosts and antifields. The ghost spectrum for the model
under study comprises the tensor fields 
\begin{equation}
\left( \stackrel{(m)}{\eta }_{\mu _{1}\cdots \mu _{k}|\nu _{1}\cdots \nu
_{k-m-1}}\right) _{m=\overline{0,k-1}},  \label{k53}
\end{equation}
with the Grassmann parities 
\begin{equation}
\varepsilon \left( \stackrel{(m)}{\eta }_{\mu _{1}\cdots \mu _{k}|\nu
_{1}\cdots \nu _{k-m-1}}\right) =\left( m+1\right) \;\mathrm{mod}\;2,\;m=%
\overline{0,k-1}.  \label{k54}
\end{equation}
The fermionic ghosts $\stackrel{(0)}{\eta }_{\mu _{1}\cdots \mu _{k}|\nu
_{1}\cdots \nu _{k-1}}$ are associated with the gauge parameters $\epsilon
_{\mu _{1}\cdots \mu _{k}|\nu _{1}\cdots \nu _{k-1}}$ from the
transformations (\ref{k8}), while the rest of the ghosts are due to the
reducibility parameters of order $m$ that appear in the relations (\ref{k14}%
) and (\ref{k16}). In order to make compatible the behaviour of the ghosts
with that of the gauge and reducibility parameters, we ask that $\left( 
\stackrel{(m)}{\eta }_{\mu _{1}\cdots \mu _{k}|\nu _{1}\cdots \nu
_{k-m-1}}\right) $ for $m=\overline{0,k-2}$ display the mixed symmetry $%
\left( k,k-m-1\right) $ of the two-column Young diagrams (\ref{k8a}) and (%
\ref{k20}), so they are separately antisymmetric in the first $k$ and
respectively last $\left( k-m-1\right) $ indices and satisfy the identities 
\begin{equation}
\stackrel{(m)}{\eta }_{\left[ \mu _{1}\cdots \mu _{k}|\nu _{1}\right] \nu
_{2}\cdots \nu _{k-m-1}}\equiv 0,\;m=\overline{0,k-2}  \label{k55}
\end{equation}
while the ghost for ghost $\stackrel{(k-1)}{\eta }_{\mu _{1}\cdots \mu _{k}}$
is completely antisymmetric. The antifield spectrum is organized into the
antifields 
\begin{equation}
t^{*\mu _{1}\cdots \mu _{k}|\nu _{1}\cdots \nu _{k}},\;\left( \stackrel{(m)}{%
\eta }^{*\mu _{1}\cdots \mu _{k}|\nu _{1}\cdots \nu _{k-m-1}}\right) _{m=%
\overline{0,k-1}}  \label{k56}
\end{equation}
corresponding to the original tensor field and to the ghosts, of statistics
opposite to that of the associated fields/ghosts 
\[
\varepsilon \left( t^{*\mu _{1}\cdots \mu _{k}|\nu _{1}\cdots \nu
_{k}}\right) =1,\;\varepsilon \left( \stackrel{(m)}{\eta }^{*\mu _{1}\cdots
\mu _{k}|\nu _{1}\cdots \nu _{k-m-1}}\right) =m\;\mathrm{mod}\;2. 
\]
Obviously, the antifields exhibit the same mixed properties like the
associated field/ghosts, so they are separately antisymmetric in the first $%
k $ and respectively last $k$ or $\left( k-m-1\right) $ indices and satisfy
the identities 
\begin{equation}
t^{*\left[ \mu _{1}\cdots \mu _{k}|\nu _{1}\right] \nu _{2}\cdots \nu
_{k}}\equiv 0,\;\stackrel{(m)}{\eta }^{*\left[ \mu _{1}\cdots \mu _{k}|\nu
_{1}\right] \nu _{2}\cdots \nu _{k-m-1}}\equiv 0,\;m=\overline{0,k-2},
\label{k57}
\end{equation}
while $\stackrel{(k-1)}{\eta }^{*\mu _{1}\cdots \mu _{k}}$ is completely
antisymmetric. In addition, $t^{*\mu _{1}\cdots \mu _{k}|\nu _{1}\cdots \nu
_{k}}$ is symmetric under the inter-change between the two sets of indices 
\begin{equation}
t^{*\mu _{1}\cdots \mu _{k}|\nu _{1}\cdots \nu _{k}}=t^{*\nu _{1}\cdots \nu
_{k}|\mu _{1}\cdots \mu _{k}}.  \label{k58}
\end{equation}
We will denote the various traces of $t^{*\mu _{1}\cdots \mu _{k}|\nu
_{1}\cdots \nu _{k}}$ by 
\begin{equation}
t^{*\mu _{m+1}\cdots \mu _{k}|\nu _{m+1}\nu _{2}\cdots \nu _{k}}=\sigma
_{\mu _{1}\nu _{1}}\cdots \sigma _{\mu _{m}\nu _{m}}t^{*\mu _{1}\cdots \mu
_{k}|\nu _{1}\nu _{2}\cdots \nu _{k}},\;m=\overline{1,k},  \label{k59}
\end{equation}
being understood that we maintain the conventions (\ref{conv}).

As both the gauge generators and reducibility functions for this model are
field-independent, it follows that the associated BRST differential ($%
s^{2}=0 $) splits into 
\begin{equation}
s=\delta +\gamma ,  \label{k60}
\end{equation}
where $\delta $ represents the Koszul-Tate differential ($\delta ^{2}=0$),
graded by the antighost number $\mathrm{agh}$ ($\mathrm{agh}\left( \delta
\right) =-1$), while $\gamma $ stands for the exterior derivative along the
gauge orbits and turns out to be a true differential ($\gamma ^{2}=0$) that
anticommutes with $\delta $ ($\delta \gamma +\gamma \delta =0$), whose
degree is named pure ghost number $\mathrm{pgh}$ ($\mathrm{pgh}\left( \gamma
\right) =1$). These two degrees do not interfere ($\mathrm{agh}\left( \gamma
\right) =0$, $\mathrm{pgh}\left( \delta \right) =0$). The overall degree
that grades the BRST differential is known as the ghost number ($\mathrm{gh}$%
) and is defined like the difference between the pure ghost number and the
antighost number, such that $\mathrm{gh}\left( s\right) =\mathrm{gh}\left(
\delta \right) =\mathrm{gh}\left( \gamma \right) =1$. According to the
standard rules of the BRST method, the corresponding degrees of the
generators from the BRST complex are valued like 
\begin{eqnarray}
\mathrm{pgh}\left( t_{\mu _{1}\cdots \mu _{k}|\nu _{1}\cdots \nu
_{k}}\right) &=&0,\;\mathrm{pgh}\left( \stackrel{(m)}{\eta }_{\mu _{1}\cdots
\mu _{k}|\nu _{1}\cdots \nu _{k-m-1}}\right) =m+1,  \label{k61} \\
\mathrm{pgh}\left( t^{*\mu _{1}\cdots \mu _{k}|\nu _{1}\cdots \nu
_{k}}\right) &=&\mathrm{pgh}\left( \stackrel{(m)}{\eta }^{*\mu _{1}\cdots
\mu _{k}|\nu _{1}\cdots \nu _{k-m-1}}\right) =0,  \label{k62} \\
\mathrm{agh}\left( t_{\mu _{1}\cdots \mu _{k}|\nu _{1}\cdots \nu
_{k}}\right) &=&\mathrm{agh}\left( \stackrel{(m)}{\eta }_{\mu _{1}\cdots \mu
_{k}|\nu _{1}\cdots \nu _{k-m-1}}\right) =0,  \label{k63} \\
\mathrm{agh}\left( t^{*\mu _{1}\cdots \mu _{k}|\nu _{1}\cdots \nu
_{k}}\right) &=&1,\;\mathrm{agh}\left( \stackrel{(m)}{\eta }^{*\mu
_{1}\cdots \mu _{k}|\nu _{1}\cdots \nu _{k-m-1}}\right) =m+2,  \label{k64}
\end{eqnarray}
with $m=\overline{0,k-1}$, while the actions of $\delta $ and $\gamma $ on
them are given by 
\begin{equation}
\gamma t_{\mu _{1}\cdots \mu _{k}|\nu _{1}\cdots \nu _{k}}=\stackrel{(0)}{%
\eta }_{\mu _{1}\cdots \mu _{k}|\left[ \nu _{2}\cdots \nu _{k},\nu
_{1}\right] }+\stackrel{(0)}{\eta }_{\nu _{1}\cdots \nu _{k}|\left[ \mu
_{2}\cdots \mu _{k},\mu _{1}\right] },  \label{k65}
\end{equation}
\begin{eqnarray}
&&\gamma \stackrel{(m)}{\eta }_{\mu _{1}\cdots \mu _{k}|\nu _{1}\cdots \nu
_{k-m-1}}=\partial _{\left[ \mu _{1}\right. }\stackrel{(m+1)}{\eta }_{\left.
\mu _{2}\cdots \mu _{k}\right] \left[ \nu _{1}|\nu _{2}\cdots \nu
_{k-m-1}\right] }  \nonumber \\
&&+\left( -\right) ^{k+1}\left( m+2\right) \stackrel{(m+1)}{\eta }_{\mu
_{1}\cdots \mu _{k}|\left[ \nu _{2}\cdots \nu _{k-m-1},\nu _{1}\right] },\;m=%
\overline{0,k-2}  \label{k66}
\end{eqnarray}
\begin{equation}
\gamma \stackrel{(k-1)}{\eta }_{\mu _{1}\cdots \mu _{k}}=0,  \label{k67}
\end{equation}
\begin{equation}
\gamma t^{*\mu _{1}\cdots \mu _{k}|\nu _{1}\cdots \nu _{k}}=0,\;\gamma 
\stackrel{(m)}{\eta }^{*\mu _{1}\cdots \mu _{k}|\nu _{1}\cdots \nu
_{k-m-1}}=0,\;m=\overline{0,k-1},  \label{k68}
\end{equation}
\begin{equation}
\delta t_{\mu _{1}\cdots \mu _{k}|\nu _{1}\cdots \nu _{k}}=0,\;\delta 
\stackrel{(m)}{\eta }_{\mu _{1}\cdots \mu _{k}|\nu _{1}\cdots \nu
_{k-m-1}}=0,\;m=\overline{0,k-1},  \label{k69}
\end{equation}
\begin{equation}
\delta t^{*\mu _{1}\cdots \mu _{k}|\nu _{1}\cdots \nu _{k}}=-c_{1}T^{\mu
_{1}\cdots \mu _{k}|\nu _{1}\cdots \nu _{k}},  \label{k70}
\end{equation}
\begin{equation}
\delta \stackrel{(0)}{\eta }^{*\mu _{1}\cdots \mu _{k}|\nu _{1}\cdots \nu
_{k-1}}=-2k\partial _{\rho }t^{*\mu _{1}\cdots \mu _{k}|\rho \nu _{1}\cdots
\nu _{k-1}},  \label{k71}
\end{equation}
\begin{eqnarray}
\delta \stackrel{(m)}{\eta }^{*\mu _{1}\cdots \mu _{k}|\nu _{1}\cdots \nu
_{k-m-1}} &=&  \nonumber \\
\left( -\right) ^{k-m}\left( k-m\right) \left( m+2\right) \partial _{\rho }%
\stackrel{(m-1)}{\eta }^{*\mu _{1}\cdots \mu _{k}|\rho \nu _{1}\cdots \nu
_{k-m-1}},\;m &=&\overline{1,k-1},  \label{k72}
\end{eqnarray}
with $T^{\mu _{1}\cdots \mu _{k}|\nu _{1}\cdots \nu _{k}}$ resulting from (%
\ref{k23}) and both $\delta $ and $\gamma $ taken to act like right
derivations.

The antifield-BRST differential is known to admit a canonical action in a
structure named antibracket and defined by decreeing the fields/ghosts
conjugated with the corresponding antifields, $s\cdot =\left( \cdot
,S\right) $, where $\left( ,\right) $ signifies the antibracket and $S$
denotes the canonical generator of the BRST symmetry. It is a bosonic
functional of ghost number zero involving both the field/ghost and antifield
spectra, which obeys the classical master equation 
\begin{equation}
\left( S,S\right) =0.  \label{k73}
\end{equation}
The classical master equation is equivalent with the second-order nilpotency
of $s$, $s^{2}=0$, while its solution encodes the entire gauge structure of
the associated theory. Taking into account the formulas (\ref{k65}--\ref{k72}%
), as well as the actions of $\delta $ and $\gamma $ in canonical form, we
find that the complete solution to the master equation for the model under
study reads as 
\begin{eqnarray}
S &=&S^{\mathrm{L}}\left[ t_{\mu _{1}\cdots \mu _{k}|\nu _{1}\cdots \nu
_{k}}\right] +\int d^{D}x\left( t^{*\mu _{1}\cdots \mu _{k}|\nu _{1}\cdots
\nu _{k}}\left( \stackrel{(0)}{\eta }_{\mu _{1}\cdots \mu _{k}|\left[ \nu
_{2}\cdots \nu _{k},\nu _{1}\right] }\right. \right.  \nonumber \\
&&\left. +\stackrel{(0)}{\eta }_{\nu _{1}\cdots \nu _{k}|\left[ \mu
_{2}\cdots \mu _{k},\mu _{1}\right] }\right)  \nonumber \\
&&+\sum_{m=0}^{k-2}\stackrel{(m)}{\eta }^{*\mu _{1}\cdots \mu _{k}|\nu
_{1}\cdots \nu _{k-m-1}}\left( \partial _{\left[ \mu _{1}\right. }\stackrel{%
(m+1)}{\eta }_{\left. \mu _{2}\cdots \mu _{k}\right] \left[ \nu _{1}|\nu
_{2}\cdots \nu _{k-m-1}\right] }\right.  \nonumber \\
&&\left. \left. +\left( -\right) ^{k+1}\left( m+2\right) \stackrel{(m+1)}{%
\eta }_{\mu _{1}\cdots \mu _{k}|\left[ \nu _{2}\cdots \nu _{k-m-1},\nu
_{1}\right] }\right) \right) .  \label{k74}
\end{eqnarray}
The main ingredients of the antifield-BRST symmetry derived in this section
will be useful in the sequel at the analysis of the BRST cohomology for the
free, massless tensor field $\left( k,k\right) $.

\subsection{Cohomology of the exterior longitudinal derivative and related
matters\label{4.2}}

The main aim of this paper is to study of the local cohomology $H\left(
s|d\right) $ in form degree $D$ ($D\geq 2k+1$). As it will be further seen,
an indispensable ingredient in the computation of $H\left( s|d\right) $ is
the cohomology algebra of the exterior longitudinal derivative ($H\left(
\gamma \right) $). It is defined by the equivalence classes of $\gamma $%
-closed non-integrated densities $a$ of fields, ghosts, antifields and their
spacetime derivatives, $\gamma a=0$, modulo $\gamma $-exact terms. If $a\in
H\left( \gamma \right) $ is $\gamma $-exact, $a=\gamma b$, then $a$ belongs
to the class of the element zero and we call it $\gamma $-trivial. In other
words, the solution to the equation $\gamma a=0$ is unique up to $\gamma $%
-trivial objects, $a\rightarrow a+\gamma b$. The cohomology algebra $H\left(
\gamma \right) $ inherits a natural grading $H\left( \gamma \right)
=\bigoplus\nolimits_{l\geq 0}H^{l}\left( \gamma \right) $, where $l$ is the
pure ghost number. Let $a$ be an element of $H\left( \gamma \right) $ with
definite pure ghost number, antighost number and form degree ($\deg $) 
\begin{equation}
\gamma a=0,\;\mathrm{pgh}\left( a\right) =l\geq 0,\;\mathrm{agh}\left(
a\right) =j\geq 0,\;\deg \left( a\right) =p\leq D.  \label{k75}
\end{equation}
In the sequel we analyze the general form of $a$ with the above properties
with the help of the definitions (\ref{k65}--\ref{k68}).

The formula (\ref{k68}) shows that all the antifields 
\begin{equation}
\chi ^{*\Delta }\equiv \left( t^{*\mu _{1}\cdots \mu _{k}|\nu _{1}\cdots \nu
_{k}},\left( \stackrel{(m)}{\eta }^{*\mu _{1}\cdots \mu _{k}|\nu _{1}\cdots
\nu _{k-m-1}}\right) _{m=\overline{0,k-1}}\right) ,  \label{k76}
\end{equation}
and their spacetime derivatives belong (non-trivially) to $H^{0}\left(
\gamma \right) $. From (\ref{k65}) we observe that $\gamma $ acts on the
original tensor field through a gauge transformation (\ref{k8}) with the
gauge parameters replaced by the ghosts, such that the $\gamma $-closed
quantities constructed out of the tensor gauge field $t_{\mu _{1}\cdots \mu
_{k}|\nu _{1}\cdots \nu _{k}}$ and its space-time derivatives are nothing
but the gauge-invariant objects of the theory (\ref{k8}). As it was
discussed in Section 3, the only such objects are the curvature tensor (\ref
{k27}) $F_{\mu _{1}\cdots \mu _{k+1}|\nu _{1}\cdots \nu _{k+1}}$ and its
derivatives, and thus they all belong to $H^{0}\left( \gamma \right) $. With
the help of the definitions in (\ref{k66}) one prove the following theorem.

\begin{theorem}
\label{trivgama}The cohomology spaces $H^{l}\left( \gamma \right) $ of the
exterior longitudinal derivative in pure ghost number $1\leq l\leq k-1$ for
the free, massless tensor field $t_{\mu _{1}\cdots \mu _{k}|\nu _{1}\cdots
\nu _{k}}$ are all vanishing 
\begin{equation}
H^{l}\left( \gamma \right) =0,\;1\leq l\leq k-1.  \label{k77}
\end{equation}
\end{theorem}

\emph{Proof} The proof is purely computational and essentially relies on the
fact that the most general $\gamma $ -closed quantities that are linear in
the ghosts $\left( \stackrel{(m)}{\eta }_{\mu _{1}\cdots \mu _{k}|\nu
_{1}\cdots \nu _{k-m-1}}\right) _{m=\overline{0,k-2}}$ are also $\gamma $%
-exact. Since the general case is intricate and yet not illuminating, we
give below the detailed proof in the case $k=3$. The ghost spectrum
comprises the ghosts $\stackrel{(0)}{\eta }_{\mu \nu \rho |\alpha \beta }$, $%
\stackrel{(1)}{\eta }_{\mu \nu \rho |\alpha }$ and $\stackrel{(2)}{\eta }%
_{\mu \nu \rho }$, on which $\gamma $ acts through 
\begin{eqnarray}
\gamma t_{\mu \nu \rho |\alpha \beta \gamma } &=&\stackrel{(0)}{\eta }_{\mu
\nu \rho |\left[ \beta \gamma ,\alpha \right] }+\stackrel{(0)}{\eta }%
_{\alpha \beta \gamma |\left[ \nu \rho ,\mu \right] },  \label{k78} \\
\gamma \stackrel{(0)}{\eta }_{\mu \nu \rho |\alpha \beta } &=&\partial
_{\left[ \mu \right. }\stackrel{(1)}{\eta }_{\left. \nu \rho \right] \left[
\alpha |\beta \right] }+2\stackrel{(1)}{\eta }_{\mu \nu \rho |\left[ \beta
,\alpha \right] },  \label{k79} \\
\gamma \stackrel{(1)}{\eta }_{\mu \nu \rho |\alpha } &=&\partial _{\left[
\mu \right. }\stackrel{(2)}{\eta }_{\left. \nu \rho \right] \alpha
}+3\partial _{\alpha }\stackrel{(2)}{\eta }_{\mu \nu \rho },  \label{k80} \\
\gamma \stackrel{(2)}{\eta }_{\mu \nu \rho } &=&0.  \label{k81}
\end{eqnarray}
Using the formula (\ref{k79}), we notice that there is no $\gamma $-closed
linear combination of the undifferentiated ghosts of pure ghost number one $%
\stackrel{(0)}{\eta }_{\mu \nu \rho |\alpha \beta }$. Next, we analyze the
presence of $\gamma $-closed linear combinations involving the first-order
derivatives of the pure ghost number one ghosts. Taking into account the
identity $\stackrel{(0)}{\eta }_{\left[ \mu \nu \rho |\alpha \right] \beta
}\equiv 0$, it follows that the general expression of the linear combination
involving the first-order derivatives of the pure ghost number one ghosts
contains 30 real constants and reads as 
\begin{eqnarray}
A_{\mu _{1}\cdots \mu _{6}} &=&\partial _{\mu _{1}}\left( k_{1}\stackrel{(0)%
}{\eta }_{\mu _{2}\mu _{4}\mu _{6}|\mu _{3}\mu _{5}}+k_{2}\stackrel{(0)}{%
\eta }_{\mu _{3}\mu _{4}\mu _{6}|\mu _{2}\mu _{5}}+k_{3}\stackrel{(0)}{\eta }%
_{\mu _{2}\mu _{5}\mu _{6}|\mu _{3}\mu _{4}}\right.  \nonumber \\
&&\left. +k_{4}\stackrel{(0)}{\eta }_{\mu _{3}\mu _{5}\mu _{6}|\mu _{2}\mu
_{4}}+k_{5}\stackrel{(0)}{\eta }_{\mu _{4}\mu _{5}\mu _{6}|\mu _{2}\mu
_{3}}\right) +\partial _{\mu _{2}}\left( k_{6}\stackrel{(0)}{\eta }_{\mu
_{1}\mu _{4}\mu _{6}|\mu _{3}\mu _{5}}\right.  \nonumber \\
&&+k_{7}\stackrel{(0)}{\eta }_{\mu _{3}\mu _{4}\mu _{6}|\mu _{1}\mu
_{5}}+k_{8}\stackrel{(0)}{\eta }_{\mu _{1}\mu _{5}\mu _{6}|\mu _{3}\mu
_{4}}+k_{9}\stackrel{(0)}{\eta }_{\mu _{3}\mu _{5}\mu _{6}|\mu _{1}\mu _{4}}
\nonumber \\
&&\left. +k_{10}\stackrel{(0)}{\eta }_{\mu _{4}\mu _{5}\mu _{6}|\mu _{1}\mu
_{3}}\right) +\partial _{\mu _{3}}\left( k_{11}\stackrel{(0)}{\eta }_{\mu
_{1}\mu _{4}\mu _{6}|\mu _{2}\mu _{5}}+k_{12}\stackrel{(0)}{\eta }_{\mu
_{2}\mu _{4}\mu _{6}|\mu _{1}\mu _{5}}\right.  \nonumber \\
&&\left. +k_{13}\stackrel{(0)}{\eta }_{\mu _{1}\mu _{5}\mu _{6}|\mu _{2}\mu
_{4}}+k_{14}\stackrel{(0)}{\eta }_{\mu _{2}\mu _{5}\mu _{6}|\mu _{1}\mu
_{4}}+k_{15}\stackrel{(0)}{\eta }_{\mu _{4}\mu _{5}\mu _{6}|\mu _{1}\mu
_{2}}\right)  \nonumber \\
&&+\partial _{\mu _{4}}\left( k_{16}\stackrel{(0)}{\eta }_{\mu _{1}\mu
_{3}\mu _{6}|\mu _{2}\mu _{5}}+k_{17}\stackrel{(0)}{\eta }_{\mu _{2}\mu
_{3}\mu _{6}|\mu _{1}\mu _{5}}+k_{18}\stackrel{(0)}{\eta }_{\mu _{1}\mu
_{5}\mu _{6}|\mu _{2}\mu _{3}}\right.  \nonumber \\
&&\left. +k_{19}\stackrel{(0)}{\eta }_{\mu _{2}\mu _{5}\mu _{6}|\mu _{1}\mu
_{3}}+k_{20}\stackrel{(0)}{\eta }_{\mu _{3}\mu _{5}\mu _{6}|\mu _{1}\mu
_{2}}\right) +\partial _{\mu _{5}}\left( k_{21}\stackrel{(0)}{\eta }_{\mu
_{1}\mu _{3}\mu _{6}|\mu _{2}\mu _{4}}\right.  \nonumber \\
&&+k_{22}\stackrel{(0)}{\eta }_{\mu _{2}\mu _{3}\mu _{6}|\mu _{1}\mu
_{4}}+k_{23}\stackrel{(0)}{\eta }_{\mu _{1}\mu _{4}\mu _{6}|\mu _{2}\mu
_{3}}+k_{24}\stackrel{(0)}{\eta }_{\mu _{2}\mu _{4}\mu _{6}|\mu _{1}\mu _{3}}
\nonumber \\
&&\left. +k_{25}\stackrel{(0)}{\eta }_{\mu _{3}\mu _{4}\mu _{6}|\mu _{1}\mu
_{2}}\right) +\partial _{\mu _{6}}\left( k_{26}\stackrel{(0)}{\eta }_{\mu
_{1}\mu _{3}\mu _{5}|\mu _{2}\mu _{4}}+k_{27}\stackrel{(0)}{\eta }_{\mu
_{2}\mu _{3}\mu _{5}|\mu _{1}\mu _{4}}\right.  \nonumber \\
&&\left. +k_{28}\stackrel{(0)}{\eta }_{\mu _{1}\mu _{4}\mu _{5}|\mu _{2}\mu
_{3}}+k_{29}\stackrel{(0)}{\eta }_{\mu _{2}\mu _{4}\mu _{5}|\mu _{1}\mu
_{3}}+k_{30}\stackrel{(0)}{\eta }_{\mu _{3}\mu _{4}\mu _{5}|\mu _{1}\mu
_{2}}\right) .  \label{k82}
\end{eqnarray}
Requiring that $\gamma A_{\mu _{1}\cdots \mu _{6}}=0$, we find a homogeneous
algebraic system of 50 equations with 30 variables, whose rank is equal to
28. The solution to this system, expressed in terms of the two independent
constants, taken for instance to be $k_{29}$ and $k_{30}$, is 
\begin{eqnarray}
k_{1} &=&k_{2}=k_{4}=k_{6}=k_{7}=k_{11}=0,  \label{k83} \\
k_{12} &=&k_{13}=k_{16}=k_{17}=k_{22}=k_{27}=0,  \label{k84} \\
k_{9} &=&k_{14}=k_{21}=-k_{3}=-k_{8}=-k_{26}=k_{29}+k_{30},  \label{k85} \\
k_{5} &=&k_{15}=k_{18}=k_{20}=k_{24}=-k_{29},  \label{k86} \\
k_{10} &=&k_{19}=k_{23}=k_{25}=-k_{28}=-k_{30},  \label{k87}
\end{eqnarray}
so we find that 
\begin{equation}
A_{\mu _{1}\cdots \mu _{6}}=-\left( \left( k_{29}+k_{30}\right) A_{\mu
_{1}\mu _{3}\mu _{4}|\mu _{2}\mu _{5}\mu _{6}}+k_{29}A_{\mu _{1}\mu _{2}\mu
_{3}|\mu _{4}\mu _{5}\mu _{6}}\right) ,  \label{k88}
\end{equation}
where $A_{\mu _{1}\mu _{2}\mu _{3}|\mu _{4}\mu _{5}\mu _{6}}$ has the mixed
symmetry $\left( 3,3\right) $ and is given by 
\begin{equation}
A_{\mu _{1}\mu _{2}\mu _{3}|\mu _{4}\mu _{5}\mu _{6}}=\stackrel{(0)}{\eta }%
_{\mu _{1}\mu _{2}\mu _{3}|\left[ \mu _{5}\mu _{6},\mu _{4}\right] }+%
\stackrel{(0)}{\eta }_{\mu _{4}\mu _{5}\mu _{6}|\left[ \mu _{2}\mu _{3},\mu
_{1}\right] }.  \label{k89}
\end{equation}
It is easy to see from formula (\ref{k78}) that $A_{\mu _{1}\mu _{2}\mu
_{3}|\mu _{4}\mu _{5}\mu _{6}}$ is $\gamma $-exact 
\begin{equation}
A_{\mu _{1}\mu _{2}\mu _{3}|\mu _{4}\mu _{5}\mu _{6}}=\gamma \left( t_{\mu
_{1}\mu _{2}\mu _{3}|\mu _{4}\mu _{5}\mu _{6}}\right) ,  \label{k90}
\end{equation}
and thus it (and also $A_{\mu _{1}\cdots \mu _{6}}$) must be discarded from $%
H^{1}\left( \gamma \right) $ as being trivial. Along the same line, one can
prove that the only $\gamma $-closed combinations with $N\geq 2$ spacetime
derivatives of the ghosts $\stackrel{(0)}{\eta }_{\mu \nu \rho |\alpha \beta
}$ are actually polynomials with $\left( N-1\right) $ derivatives in the
elements $A_{\mu _{1}\mu _{2}\mu _{3}|\mu _{4}\mu _{5}\mu _{6}}$, which, by
means of (\ref{k90}), are $\gamma $-exact, and hence trivial in $H^{1}\left(
\gamma \right) $. In conclusion, there is no non-trivial object constructed
out of the ghosts $\stackrel{(0)}{\eta }_{\mu \nu \rho |\alpha \beta }$ and
their derivatives in $H^{1}\left( \gamma \right) $, which implies that 
\begin{equation}
H^{1}\left( \gamma \right) =0\;\mathrm{for}\;k=3,  \label{h1}
\end{equation}
as there are no other ghosts of pure ghost number equal to one in the BRST
complex.

With the help of the definition (\ref{k80}), we notice that there is no $%
\gamma $-closed linear combination of the undifferentiated ghosts of pure
ghost number two $\stackrel{(1)}{\eta }_{\mu \nu \rho |\alpha }$. Now, we
pass to the determination of $\gamma $-closed linear combinations involving
the first-order derivatives of the pure ghost number two ghosts. By means of
the identity $\stackrel{(1)}{\eta }_{\left[ \mu \nu \rho |\alpha \right]
}\equiv 0$, we get that the general expression of the linear combination
involving the first-order derivatives of the pure ghost number two ghosts
contains 15 real constants and has the form 
\begin{eqnarray}
B_{\mu _{1}\cdots \mu _{5}} &=&\partial _{\mu _{1}}\left( m_{1}\stackrel{(1)%
}{\eta }_{\mu _{2}\mu _{3}\mu _{4}|\mu _{5}}+m_{2}\stackrel{(1)}{\eta }_{\mu
_{2}\mu _{3}\mu _{5}|\mu _{4}}+m_{3}\stackrel{(1)}{\eta }_{\mu _{2}\mu
_{4}\mu _{5}|\mu _{3}}\right)  \nonumber \\
&&+\partial _{\mu _{2}}\left( m_{4}\stackrel{(1)}{\eta }_{\mu _{1}\mu
_{3}\mu _{4}|\mu _{5}}+m_{5}\stackrel{(1)}{\eta }_{\mu _{1}\mu _{3}\mu
_{5}|\mu _{4}}+m_{6}\stackrel{(1)}{\eta }_{\mu _{1}\mu _{4}\mu _{5}|\mu
_{3}}\right)  \nonumber \\
&&+\partial _{\mu _{3}}\left( m_{7}\stackrel{(1)}{\eta }_{\mu _{1}\mu
_{2}\mu _{4}|\mu _{5}}+m_{8}\stackrel{(1)}{\eta }_{\mu _{1}\mu _{2}\mu
_{5}|\mu _{4}}+m_{9}\stackrel{(1)}{\eta }_{\mu _{1}\mu _{4}\mu _{5}|\mu
_{2}}\right)  \nonumber \\
&&+\partial _{\mu _{4}}\left( m_{10}\stackrel{(1)}{\eta }_{\mu _{1}\mu
_{2}\mu _{3}|\mu _{5}}+m_{11}\stackrel{(1)}{\eta }_{\mu _{1}\mu _{2}\mu
_{5}|\mu _{3}}+m_{12}\stackrel{(1)}{\eta }_{\mu _{1}\mu _{3}\mu _{5}|\mu
_{2}}\right)  \nonumber \\
&&+\partial _{\mu _{5}}\left( m_{13}\stackrel{(1)}{\eta }_{\mu _{1}\mu
_{2}\mu _{3}|\mu _{4}}+m_{14}\stackrel{(1)}{\eta }_{\mu _{1}\mu _{2}\mu
_{4}|\mu _{3}}+m_{15}\stackrel{(1)}{\eta }_{\mu _{1}\mu _{3}\mu _{4}|\mu
_{2}}\right) .  \label{k91}
\end{eqnarray}
The requirement $\gamma B_{\mu _{1}\cdots \mu _{5}}=0$ leads to a
homogeneous algebraic system of 10 equations with 15 variables, whose rank
is equal to 5. The solution to this system, expressed in terms of the five
independent constants, taken for instance to be $\left( m_{i}\right) _{i=%
\overline{11,15}}$ is 
\begin{eqnarray}
m_{1} &=&-\frac{2m_{11}}{3}-\frac{m_{12}}{3}-\frac{m_{13}}{2}+\frac{5m_{14}}{%
6}+\frac{m_{15}}{6},  \label{k92} \\
m_{2} &=&m_{11}+\frac{m_{13}}{2}-\frac{m_{14}}{2}-\frac{m_{15}}{2},
\label{k93} \\
m_{3} &=&-m_{11}-m_{12}+m_{14}+m_{15},  \label{k94} \\
m_{4} &=&\frac{2m_{11}}{3}+\frac{m_{12}}{3}+\frac{m_{13}}{2}-\frac{5m_{14}}{6%
}-\frac{7m_{15}}{6},  \label{k95} \\
m_{5} &=&-m_{11}-m_{12}-\frac{m_{13}}{2}+\frac{m_{14}}{2}+\frac{m_{15}}{2},
\label{k96} \\
m_{6} &=&m_{11}-m_{14},\;m_{9}=m_{12}-m_{15},  \label{k97} \\
m_{7} &=&\frac{m_{11}}{3}+\frac{2m_{12}}{3}-\frac{m_{13}}{2}-\frac{7m_{14}}{6%
}-\frac{5m_{15}}{6},  \label{k98} \\
m_{8} &=&-m_{11}-m_{12}+\frac{m_{13}}{2}+\frac{m_{14}}{2}+\frac{m_{15}}{2},
\label{k99} \\
m_{10} &=&-\frac{m_{11}}{3}+\frac{m_{12}}{3}-m_{13}-\frac{m_{14}}{3}+\frac{%
m_{15}}{3},  \label{k100}
\end{eqnarray}
which further yields 
\begin{eqnarray}
B_{\mu _{1}\cdots \mu _{5}} &=&\frac{1}{3}\left( 2m_{14}-m_{11}\right)
B_{\mu _{1}\mu _{2}\mu _{5}|\mu _{3}\mu _{4}}+\frac{1}{6}\left(
7m_{14}-2m_{11}\right) B_{\mu _{3}\mu _{4}\mu _{5}|\mu _{1}\mu _{2}} 
\nonumber \\
&&+\frac{1}{6}\left( m_{15}-2m_{12}\right) B_{\mu _{1}\mu _{3}\mu _{5}|\mu
_{2}\mu _{4}}+\frac{1}{6}\left( m_{15}-2m_{12}\right) B_{\mu _{2}\mu _{4}\mu
_{5}|\mu _{1}\mu _{3}}  \nonumber \\
&&-\frac{m_{13}}{2}B_{\mu _{1}\mu _{2}\mu _{3}|\mu _{4}\mu _{5}}-\frac{m_{14}%
}{2}B_{\mu _{1}\mu _{2}\mu _{4}|\mu _{3}\mu _{5}}-\frac{m_{15}}{2}B_{\mu
_{1}\mu _{3}\mu _{4}|\mu _{2}\mu _{5}},  \label{k101}
\end{eqnarray}
where $B_{\mu _{1}\mu _{2}\mu _{3}|\mu _{4}\mu _{5}}$ has the mixed symmetry 
$\left( 3,2\right) $ and reads as 
\begin{equation}
B_{\mu _{1}\mu _{2}\mu _{3}|\mu _{4}\mu _{5}}=\partial _{\left[ \mu
_{1}\right. }\stackrel{(1)}{\eta }_{\left. \mu _{2}\mu _{3}\right] \left[
\mu _{4}|\mu _{5}\right] }+2\stackrel{(1)}{\eta }_{\mu _{1}\mu _{2}\mu
_{3}|\left[ \mu _{5},\mu _{4}\right] }.  \label{k102}
\end{equation}
On account of the definition (\ref{k79}) it results that $B_{\mu _{1}\mu
_{2}\mu _{3}|\mu _{4}\mu _{5}}$ is $\gamma $-exact 
\begin{equation}
B_{\mu _{1}\mu _{2}\mu _{3}|\mu _{4}\mu _{5}}=\gamma \left( \stackrel{(0)}{%
\eta }_{\mu _{1}\mu _{2}\mu _{3}|\mu _{4}\mu _{5}}\right) ,  \label{k103}
\end{equation}
and thus it (as well as $B_{\mu _{1}\mu _{2}\mu _{3}|\mu _{4}\mu _{5}}$)
must be thrown out from $H^{2}\left( \gamma \right) $ as being trivial. In
the meantime, it can be shown that the only $\gamma $-closed combinations
with $N\geq 2$ spacetime derivatives of the ghosts $\stackrel{(1)}{\eta }%
_{\mu \nu \rho |\alpha }$ are actually polynomials with $\left( N-1\right) $
derivatives in the elements $B_{\mu _{1}\mu _{2}\mu _{3}|\mu _{4}\mu _{5}}$,
which are $\gamma $-exact due to (\ref{k103}), and hence $\gamma $-trivial.
This result allows us to state that there is no non-trivial object
constructed out of the ghosts $\stackrel{(1)}{\eta }_{\mu \nu \rho |\alpha }$
and their derivatives in $H^{2}\left( \gamma \right) $, so we have that 
\begin{equation}
H^{2}\left( \gamma \right) =0\;\mathrm{for}\;k=3,  \label{k104}
\end{equation}
on behalf of (\ref{h1}) and since there are no other ghosts of pure ghost
number equal to two in the BRST complex. $\blacksquare $

Finally, we investigate the definition (\ref{k67}). It shows that the
undifferentiated ghosts of pure ghost number equal to $k$, $\stackrel{(k-1)}{%
\eta }_{\mu _{1}\cdots \mu _{k}}$, belong to $H\left( \gamma \right) $. The $%
\gamma $- closedness of $\stackrel{(k-1)}{\eta }_{\mu _{1}\cdots \mu _{k}}$
further implies that all their derivatives are also $\gamma $-closed.
Regarding their first-order derivatives, from the definition (\ref{k66}) for 
$m=k-2$ we observe that their symmetric part is $\gamma $-exact 
\begin{equation}
\partial _{\left( \mu _{1}\right. }\stackrel{(k-1)}{\eta }_{\left. \mu
_{2}\right) \mu _{3}\cdots \mu _{k+1}}\equiv \gamma \left( \frac{1}{k+1}%
\stackrel{(k-2)}{\eta }_{\mu _{3}\cdots \mu _{k+1}\left( \mu _{1}|\mu
_{2}\right) }\right) ,  \label{k105}
\end{equation}
where $\left( \mu \nu \cdots \right) $ denotes plain symmetrization with
respect to the indices between brackets without normalization factors, such
that $\partial _{\left( \mu _{1}\right. }\stackrel{(k-1)}{\eta }_{\left. \mu
_{2}\right) \mu _{3}\cdots \mu _{k+1}}$ will be removed from $H\left( \gamma
\right) $. Meanwhile, their complete antisymmetric part $\partial _{\left[
\mu _{1}\right. }\stackrel{(k-1)}{\eta }_{\left. \mu _{2}\cdots \mu
_{k+1}\right] }$ is not $\gamma $-exact, and hence can be taken as a
non-trivial representative of $H^{k}\left( \gamma \right) $. Actually, due
to the relations 
\begin{eqnarray}
\partial _{\mu _{1}}\stackrel{(k-1)}{\eta }_{\mu _{2}\cdots \mu _{k+1}} &=&%
\frac{1}{k+1}\partial _{\left[ \mu _{1}\right. }\stackrel{(k-1)}{\eta }%
_{\left. \mu _{2}\cdots \mu _{k+1}\right] }  \nonumber \\
&&+\gamma \left( \frac{\left( -\right) ^{k+1}}{k+1}\stackrel{(k-2)}{\eta }%
_{\mu _{2}\cdots \mu _{k+1}|\mu _{1}}\right) ,  \label{k105a}
\end{eqnarray}
\begin{eqnarray}
&&\frac{1}{2}\partial _{\left[ \mu _{1}\right. }\stackrel{(k-1)}{\eta }%
_{\left. \mu _{2}\right] \mu _{3}\cdots \mu _{k+1}}=\frac{1}{k+1}\partial
_{\left[ \mu _{1}\right. }\stackrel{(k-1)}{\eta }_{\left. \mu _{2}\cdots \mu
_{k+1}\right] }  \nonumber \\
&&+\gamma \left[ -\frac{1}{k+1}\left( \left( -\right) ^{k}\stackrel{(k-2)}{%
\eta }_{\mu _{2}\cdots \mu _{k+1}|\mu _{1}}+\frac{1}{2}\stackrel{(k-2)}{\eta 
}_{\mu _{3}\cdots \mu _{k+1}\left( \mu _{1}|\mu _{2}\right) }\right) \right]
,  \label{k105b}
\end{eqnarray}
it is clear that $\partial _{\mu _{1}}\stackrel{(k-1)}{\eta }_{\mu
_{2}\cdots \mu _{k+1}}$ and $\frac{1}{2}\partial _{\left[ \mu _{1}\right. }%
\stackrel{(k-1)}{\eta }_{\left. \mu _{2}\right] \mu _{3}\cdots \mu _{k+1}}$
are in the same equivalence class from $H^{k}\left( \gamma \right) $%
\begin{eqnarray}
\partial _{\mu _{1}}\stackrel{(k-1)}{\eta }_{\mu _{2}\cdots \mu _{k+1}}
&\sim &\frac{1}{2}\partial _{\left[ \mu _{1}\right. }\stackrel{(k-1)}{\eta }%
_{\left. \mu _{2}\right] \mu _{3}\cdots \mu _{k+1}}  \nonumber \\
&\sim &\frac{1}{k+1}\partial _{\left[ \mu _{1}\right. }\stackrel{(k-1)}{\eta 
}_{\left. \mu _{2}\cdots \mu _{k+1}\right] }.  \label{k105c}
\end{eqnarray}
As a consequence, for $\mu _{m}\neq \mu _{n}$ with $m,n=\overline{1,\cdots
k+1}$ any of them can be used as a non-trivial representative of $%
H^{k}\left( \gamma \right) $. After some calculations, we find that all the
second-order derivatives of the ghosts $\stackrel{(k-1)}{\eta }_{\mu
_{1}\cdots \mu _{k}}$ are $\gamma $-exact 
\begin{eqnarray}
\partial _{\nu _{1}}\partial _{\nu _{2}}\stackrel{(k-1)}{\eta }_{\mu
_{1}\cdots \mu _{k}} &=&-\frac{1}{2k\left( k+1\right) }\gamma \left(
\partial _{\left[ \mu _{1}\right. }\stackrel{(k-2)}{\eta }_{\left. \mu
_{2}\cdots \mu _{k}\right] \,\left( \nu _{1}|\nu _{2}\right) }\right. 
\nonumber \\
&&\left. +\left( -\right) ^{k}\left( k+1\right) \stackrel{(k-2)}{\eta }_{\mu
_{1}\cdots \mu _{k}|\left( \nu _{1},\nu _{2}\right) }\right) ,  \label{k106}
\end{eqnarray}
and so will be their higher-order derivatives, such that they all disappear
from $H\left( \gamma \right) $. In conclusion, the only non-trivial
combinations in $H\left( \gamma \right) $ constructed from the ghosts of
pure ghost number equal to $k$ are polynomials in $\stackrel{(k-1)}{\eta }%
_{\mu _{1}\cdots \mu _{k}}$ and $\partial _{\left[ \mu _{1}\right. }%
\stackrel{(k-1)}{\eta }_{\left. \mu _{2}\cdots \mu _{k+1}\right] }$.
Combining this result with the previous one on $H^{0}\left( \gamma \right) $
being non-vanishing and with (\ref{k77}) , we have actually proved that only
the cohomological spaces $H^{kl}\left( \gamma \right) $ with $l\geq 0$ are
non-vanishing for the model under consideration or, equivalently, that 
\begin{equation}
H^{l^{\prime }}\left( \gamma \right) =0,\;\mathrm{for\;all}\;l^{\prime }\neq
kl.  \label{k106a}
\end{equation}

According to the results exposed so far, we can state that the \emph{general
local solution} to the equation (\ref{k75}) for $\mathrm{pgh}\left( a\right)
=kl>0$ is, up to trivial, $\gamma $-exact contributions, of the type 
\begin{equation}
a=\sum_{J}\alpha _{J}\left( \left[ \chi ^{*\Delta }\right] ,\left[ F_{\mu
_{1}\cdots \mu _{k+1}|\nu _{1}\cdots \nu _{k+1}}\right] \right) e^{J}\left( 
\stackrel{(k-1)}{\eta }_{\mu _{1}\cdots \mu _{k}},\partial _{\left[ \mu
_{1}\right. }\stackrel{(k-1)}{\eta }_{\left. \mu _{2}\cdots \mu
_{k+1}\right] }\right) ,  \label{r79}
\end{equation}
where the notation $f\left( \left[ q\right] \right) $ means that the
function $f$ depends on the variable $q$ and its subsequent derivatives up
to a finite number. In the above, $e^{J}$ are the elements of pure ghost
number $kl$ (and obviously of antighost number zero) of a basis in the space
of polynomials in $\stackrel{(k-1)}{\eta }_{\mu _{1}\cdots \mu _{k}}$ and $%
\partial _{\left[ \mu _{1}\right. }\stackrel{(k-1)}{\eta }_{\left. \mu
_{2}\cdots \mu _{k+1}\right] }$ 
\begin{equation}
\mathrm{pgh}\left( e^{J}\right) =kl>0,\;\mathrm{agh}\left( e^{J}\right) =0.
\label{t82}
\end{equation}
The objects $\alpha _{J}$ (obviously non-trivial in $H^{0}\left( \gamma
\right) $) were taken to have a bounded number of derivatives, and therefore
they are polynomials in the antifields $\chi ^{*\Delta }$, in the curvature
tensor $F_{\mu _{1}\cdots \mu _{k+1}|\nu _{1}\cdots \nu _{k+1}}$, as well as
in their derivatives. In agreement with (\ref{k75}), they display the
properties 
\begin{equation}
\mathrm{pgh}\left( \alpha _{J}\right) =0,\;\mathrm{agh}\left( \alpha
_{J}\right) =j\geq 0,\;\deg \left( \alpha _{J}\right) =p\leq D.  \label{t83}
\end{equation}
In the case $l=0$, the general (non-trivial) local elements of $H\left(
\gamma \right) $ are precisely $\alpha _{J}\left( \left[ \chi ^{*\Delta
}\right] ,\left[ F_{\mu _{1}\cdots \mu _{k+1}|\nu _{1}\cdots \nu
_{k+1}}\right] \right) $, which will be called ``invariant polynomials'' in
what follows. At zero antighost number, the invariant polynomials are
polynomials in the curvature tensor $F_{\mu _{1}\cdots \mu _{k+1}|\nu
_{1}\cdots \nu _{k+1}}$ and its derivatives.

In order to analyze the local cohomology $H\left( s|d\right) $ we are going
to need, besides $H\left( \gamma \right) $, also the cohomology of the
exterior spacetime differential $H\left( d\right) $ in the space of
invariant polynomials and other basic properties, which are addressed below.

\begin{theorem}
\label{hginv}The cohomology of $d$ in form degree strictly less than $D$ is
trivial in the space of invariant polynomials with strictly positive
antighost number. This means that the conditions 
\begin{equation}
\gamma \alpha =0,\;d\alpha =0,\;\mathrm{agh}\left( \alpha \right) >0,\;\deg
\alpha <D,\;\alpha =\alpha \left( \left[ \chi ^{*\Delta }\right] ,\left[
F\right] \right) ,  \label{A1}
\end{equation}
imply 
\begin{equation}
\alpha =d\beta ,  \label{A2}
\end{equation}
for some invariant polynomial $\beta \left( \left[ \chi ^{*\Delta }\right]
,\left[ F\right] \right) $.
\end{theorem}

\emph{Proof} In (\ref{A1}), the notation $F$ signifies the curvature tensor,
of components $F_{\mu _{1}\cdots \mu _{k+1}|\nu _{1}\cdots \nu _{k+1}}$, and 
$\chi ^{*\Delta }$ is explained in (\ref{k76}). Meanwhile, $\deg \alpha $ is
the form degree of $\alpha $. In order to prove the theorem, we decompose $d$
as 
\begin{equation}
d=d_{0}+d_{1},  \label{A4}
\end{equation}
where $d_{1}$ acts on the antifields $\chi ^{*\Delta }$ and their
derivatives only, while $d_{0}$ acts on the curvature tensor and its
derivatives 
\begin{equation}
d_{0}=\partial _{\mu }^{0}dx^{\mu },\;d_{1}=\partial _{\mu }^{1}dx^{\mu },
\label{A5}
\end{equation}
with 
\begin{eqnarray}
\partial _{\mu }^{0} &=&F_{\mu _{1}\cdots \mu _{k+1}|\nu _{1}\cdots \nu
_{k+1},\mu }\frac{\partial }{\partial F_{\mu _{1}\cdots \mu _{k+1}|\nu
_{1}\cdots \nu _{k+1}}}  \nonumber \\
&&+F_{\mu _{1}\cdots \mu _{k+1}|\nu _{1}\cdots \nu _{k+1},\mu \nu }\frac{%
\partial }{\partial F_{\mu _{1}\cdots \mu _{k+1}|\nu _{1}\cdots \nu
_{k+1},\nu }}+\cdots ,  \label{A6}
\end{eqnarray}
\begin{equation}
\partial _{\mu }^{1}=\chi _{\;\;\;\;,\mu }^{*\Delta }\frac{\partial ^{L}}{%
\partial \chi ^{*\Delta }}+\chi _{\;\;\;\;,\mu \nu }^{*\Delta }\frac{%
\partial ^{L}}{\partial \chi _{\;\;\;\;,\nu }^{*\Delta }}+\cdots .
\label{A7}
\end{equation}
Obviously, $d^{2}=0$ on invariant polynomials is equivalent with the
nilpotency and anticommutation of its components acting on invariant
polynomials 
\begin{equation}
d_{0}^{2}=0=d_{1}^{2},\;d_{0}d_{1}+d_{1}d_{0}=0.  \label{A8}
\end{equation}
The action of $d_{0}$ on a given invariant polynomial with say $l$
derivatives of $F$ and $m$ derivatives of $\chi ^{*\Delta }$ results in an
invariant polynomial with $\left( l+1\right) $ derivatives of $F$ and $m$
derivatives of $\chi ^{*\Delta }$, while the action of $d_{1}$ on the same
object leads to an invariant polynomial with $l$ derivatives of $F$ and $%
\left( m+1\right) $ derivatives of $\chi ^{*\Delta }$. In particular, $d_{0}$
gives zero when acting on an invariant polynomial that does not involve the
curvature or its derivatives, and the same is valid with respect to $d_{1}$
acting on an invariant polynomial that does not depend on any of the
antifields or their derivatives. From (\ref{A6}--\ref{A7}) we observe that 
\begin{equation}
\mathrm{agh}\left( d_{0}\right) =\mathrm{agh}\left( d_{1}\right) =\mathrm{agh%
}\left( d\right) =0,  \label{A8a}
\end{equation}
such that neither of them change the antighost number of the objects on
which they act.

The antifields $\chi ^{*\Delta }$ verify no relations between themselves and
their derivatives, except the usual symmetry properties of the type $\chi
_{\;\;\;\;,\mu \nu }^{*\Delta }=\chi _{\;\;\;\;,\nu \mu }^{*\Delta }$, and
accordingly will be named ``foreground'' fields. On the contrary, the
derivatives of the components of the curvature tensor satisfy the Bianchi II
identities (\ref{k31}), and in view of this we say that $F_{\mu _{1}\cdots
\mu _{k+1}|\nu _{1}\cdots \nu _{k+1}}$ are ``background'' fields. So, $d_{0}$
acts only on the background fields and their derivatives, while $d_{1}$ acts
only on the foreground fields and their derivatives. According to the
proposition on page 363 in \cite{dubplb}, we have that the entire cohomology
of $d_{1}$ in form degree strictly less than $D$ is trivial in the space of
invariant polynomials with strictly positive antighost number. This means
that 
\begin{equation}
\alpha =\alpha \left( \left[ \chi ^{*\Delta }\right] ,\left[ F\right]
\right) ,\;\mathrm{agh}\left( \alpha \right) =j>0,\;\deg \left( \alpha
\right) =p<D,\;d_{1}\alpha =0,  \label{A9}
\end{equation}
implies that 
\begin{equation}
\alpha =d_{1}\beta ,  \label{A10}
\end{equation}
with 
\begin{equation}
\beta =\beta \left( \left[ \chi ^{*\Delta }\right] ,\left[ F\right] \right)
,\;\mathrm{agh}\left( \beta \right) =j>0,\;\deg \left( \beta \right) =p-1.
\label{A11}
\end{equation}
In particular, we have that if an invariant polynomial (of form degree $p<D$
and with strictly positive antighost number) depending only on the
undifferentiated antifields is $d_{1}$-closed, then it vanishes 
\begin{eqnarray}
&&\left( \bar{\alpha}=\bar{\alpha}\left( \chi ^{*\Delta },\left[ F\right]
\right) ,\;\mathrm{agh}\left( \bar{\alpha}\right) >0,\right.  \nonumber \\
&&\left. \deg \left( \bar{\alpha}\right) =p<D,\;d_{1}\bar{\alpha}=0\right)
\Rightarrow \bar{\alpha}=0.  \label{A11a}
\end{eqnarray}
Only $d_{0}$ has non-trivial cohomology. For instance, any form depending
only on the antifields and their derivatives is $d_{0}$-closed, but it is
clearly not $d_{0}$-exact.

From now on, the proof is standard material and relies on decomposing $%
\alpha $ according to the number of derivatives of the antifields and on
using the triviality of the cohomology of $d_{1}$ in form degree strictly
less than $D$ in the space of invariant polynomials with strictly positive
antighost numbers. For further details, see \cite{hepth04}. $\blacksquare $

In form degree $D$ the Theorem \ref{hginv} is replaced with: let $\alpha
=\rho dx^{0}\wedge \cdots \wedge dx^{D-1}$ be a $d$-exact invariant
polynomial of form degree $D$ and of strictly positive antighost number, $%
\mathrm{agh}\left( \alpha \right) =j>0$, $\mathrm{\deg }\left( \alpha
\right) =D$, $\alpha =d\beta $. Then, one can take the $\left( D-1\right) $%
-form $\beta $ to be an invariant polynomial (of antighost number $j$). In
dual notations, this means that if $\rho $ with $\mathrm{agh}\left( \rho
\right) =j>0$ is an invariant polynomial whose Euler-Lagrange derivatives
are all vanishing, $\rho =\partial _{\mu }j^{\mu }$, then $j^{\mu }$ can be
taken to be also invariant. Theorem \ref{hginv} can be generalized as
follows.

\begin{theorem}
\label{hdinhg}The cohomology of $d$ computed in $H\left( \gamma \right) $ is
trivial in form degree strictly less than $D$ and in strictly positive
antighost number 
\begin{equation}
H_{p}^{g,j}\left( d,H\left( \gamma \right) \right) =0,\;j>0,\;p<D,
\label{A29}
\end{equation}
where $p$ is the form degree, $j$ is the antighost number and $g$ is the
ghost number.
\end{theorem}

\emph{Proof} The proof can be realized in a standard manner, like, for
instance, in \cite{hepth04,lingr}. It is however useful to mention that the
operator $\bar{D}$ is defined in this case through the relations 
\begin{eqnarray}
\bar{D}\alpha \left( \left[ \chi ^{*\Delta }\right] ,\left[ F\right] \right)
&=&d\alpha \left( \left[ \chi ^{*\Delta }\right] ,\left[ F\right] \right) ,
\label{A40A} \\
\bar{D}\stackrel{(k-1)}{\eta }_{\mu _{1}\cdots \mu _{k}} &=&\frac{1}{k+1}%
\partial _{\left[ \alpha \right. }\stackrel{(k-1)}{\eta }_{\left. \mu
_{1}\cdots \mu _{k}\right] }dx^{\alpha },  \label{A40B} \\
\bar{D}\left( \partial _{\left[ \alpha \right. }\stackrel{(k-1)}{\eta }%
_{\left. \mu _{1}\cdots \mu _{k}\right] }\right)  &=&0,  \label{A40C} \\
\bar{D}\left( \gamma b\right)  &=&0,  \label{A40CA}
\end{eqnarray}
which is easily seen to be a differential in $H\left( \gamma \right) $, $%
\bar{D}^{2}a=0$ for any $a$ with $\gamma a=0$. According to the relation (%
\ref{k66}) for $m=k-2$, we have that 
\begin{eqnarray}
d\stackrel{(k-1)}{\eta }_{\mu _{1}\cdots \mu _{k}} &=&\left( \partial
_{\alpha }\stackrel{(k-1)}{\eta }_{\mu _{1}\cdots \mu _{k}}\right)
dx^{\alpha }=\frac{1}{k+1}\partial _{\left[ \alpha \right. }\stackrel{(k-1)}{%
\eta }_{\left. \mu _{1}\cdots \mu _{k}\right] }dx^{\alpha }  \nonumber \\
&&+\gamma \left( \frac{\left( -\right) ^{k}}{k+1}\stackrel{(k-2)}{\eta }%
_{\mu _{1}\cdots \mu _{k}|\alpha }dx^{\alpha }\right)   \nonumber \\
&=&\bar{D}\stackrel{(k-1)}{\eta }_{\mu _{1}\cdots \mu _{k}}+\gamma \left( 
\frac{\left( -\right) ^{k}}{k+1}\stackrel{(k-2)}{\eta }_{\mu _{1}\cdots \mu
_{k}|\alpha }dx^{\alpha }\right) .  \label{A40D}
\end{eqnarray}
Moreover, from (\ref{A40B}--\ref{A40C}) we observe that 
\begin{equation}
\bar{D}e^{J}=A_{I}^{J}e^{I},  \label{A40H}
\end{equation}
for some constant matrix of elements $A_{I}^{J}$, that involves $dx^{\alpha }
$, such that $\bar{D}e^{J}$ is $\gamma $-closed, but not $\gamma $-exact 
\begin{equation}
de^{J}=\bar{D}e^{J}+\gamma \hat{e}^{J}=\sum_{I}A_{I}^{J}e^{I}+\gamma \hat{e}%
^{J},  \label{A40I}
\end{equation}
where $\hat{e}^{J}$ depends in general on $\stackrel{(k-1)}{\eta }_{\mu
_{1}\cdots \mu _{k}}$, $\partial _{\left[ \alpha \right. }\stackrel{(k-1)}{%
\eta }_{\left. \mu _{1}\cdots \mu _{k}\right] }$ and $\left[ \stackrel{(k-2)%
}{\eta }_{\mu _{1}\cdots \mu _{k}|\alpha }\right] $. Here, $e^{J}$ are the
elements with pure ghost number $kl$ ($l>0$) of a basis in the space of
polynomials in $\stackrel{(k-1)}{\eta }_{\mu _{1}\cdots \mu _{k}}$ and $%
\partial _{\left[ \alpha \right. }\stackrel{(k-1)}{\eta }_{\left. \mu
_{1}\cdots \mu _{k}\right] }$. Taking into account the relations (\ref{A40A}%
), (\ref{A40CA}) and (\ref{A40I}), we conclude that the differential $d$ on $%
H\left( \gamma \right) $ coincides with $\bar{D}$%
\begin{equation}
da=\bar{D}a+\gamma b,\;a\in H\left( \gamma \right) .  \label{dDgamma}
\end{equation}
Accordingly, from (\ref{A40A}) and (\ref{A40I}) we can write (for any
non-trivial element $a\in H^{kl}\left( \gamma \right) $, with $l>0$) that 
\begin{equation}
da=\bar{D}a+\gamma \left( \sum \alpha _{J}\hat{e}^{J}\right) ,
\label{eqnew1}
\end{equation}
where 
\begin{equation}
\bar{D}a=\sum \left( \alpha _{J}\bar{D}e^{J}\pm \left( d\alpha _{J}\right)
e^{J}\right) .  \label{eqnew2}
\end{equation}
We decompose $\bar{D}$ as a sum of two operators 
\begin{equation}
\bar{D}=\bar{D}_{0}+\bar{D}_{1},  \label{A42}
\end{equation}
defined through 
\begin{eqnarray}
\bar{D}_{0}\alpha \left( \left[ \chi ^{*\Delta }\right] ,\left[ F\right]
\right)  &=&\bar{D}\alpha \left( \left[ \chi ^{*\Delta }\right] ,\left[
F\right] \right) =d\alpha ,  \label{A43a} \\
\bar{D}_{0}\stackrel{(k-1)}{\eta }_{\mu _{1}\cdots \mu _{k}} &=&0,
\label{A43b} \\
\bar{D}_{0}\left( \partial _{\left[ \alpha \right. }\stackrel{(k-1)}{\eta }%
_{\left. \mu _{1}\cdots \mu _{k}\right] }\right)  &=&0,  \label{A43c} \\
\bar{D}_{0}\left( \gamma b\right)  &=&0,  \label{A43d} \\
\bar{D}_{1}\alpha \left( \left[ \chi ^{*\Delta }\right] ,\left[ F\right]
\right)  &=&0,  \label{A43e} \\
\bar{D}_{1}\stackrel{(k-1)}{\eta }_{\mu _{1}\cdots \mu _{k}} &=&\bar{D}%
\stackrel{(k-1)}{\eta }_{\mu _{1}\cdots \mu _{k}}=\frac{1}{k+1}\partial
_{\left[ \alpha \right. }\stackrel{(k-1)}{\eta }_{\left. \mu _{1}\cdots \mu
_{k}\right] }dx^{\alpha },  \label{A43f} \\
\bar{D}_{1}\left( \partial _{\left[ \alpha \right. }\stackrel{(k-1)}{\eta }%
_{\left. \mu _{1}\cdots \mu _{k}\right] }\right)  &=&0,  \label{A43g} \\
\bar{D}_{1}\left( \gamma b\right)  &=&0,  \label{A43h}
\end{eqnarray}
such that the nilpotency of $\bar{D}$ is equivalent to the nilpotency and
the anticommutation of its components 
\begin{equation}
\bar{D}^{2}=0\Leftrightarrow \left( \bar{D}_{0}^{2}=0=\bar{D}_{1}^{2},\;\bar{%
D}_{0}\bar{D}_{1}+\bar{D}_{1}\bar{D}_{0}=0\right) .  \label{A44}
\end{equation}
We reorganize the non-trivial elements $a\in H^{kl}\left( \gamma \right) $,
with $l>0$, like 
\begin{equation}
a=\stackrel{(0)}{a}+\stackrel{(1)}{a}+\cdots +\stackrel{(l)}{a},  \label{A45}
\end{equation}
where the piece $\left( \stackrel{(i)}{a}\right) _{i=\overline{0,l}}$
contains $i$ antisymmetrized derivatives of the ghosts $\partial _{\left[
\alpha \right. }\stackrel{(k-1)}{\eta }_{\left. \mu _{1}\cdots \mu
_{k}\right] }$ and $\left( l-i\right) $ undifferentiated ghosts $\stackrel{%
(k-1)}{\eta }_{\mu _{1}\cdots \mu _{k}}$ and call $\bar{D}$-degree the
number of factors of the type $\partial _{\left[ \alpha \right. }\stackrel{%
(k-1)}{\eta }_{\left. \mu _{1}\cdots \mu _{k}\right] }$. It is clear from (%
\ref{A43a}--\ref{A43h}) that the action of $\bar{D}_{0}$ on $\stackrel{(i)}{a%
}$ does not modify its $\bar{D}$-degree, while the action of $\bar{D}_{1}$
on the same element increases its $\bar{D}$-degree by one unit. From now the
proof of the theorem follows in a close the manner the line from \cite
{hepth04,lingr} and focuses on showing that 
\begin{equation}
H_{p}^{g,j}\left( \bar{D}\right) =0\;\mathrm{for}\;g=kl-j,\;l,j>0,\;p<D.
\label{A41a}
\end{equation}

In the meantime, we give below some common properties of $\gamma $ and $d$,
which will be employed in subsection \ref{4.4}, namely 
\begin{equation}
\gamma ^{2}=0,\;d^{2}=0,\;\gamma d+d\gamma =0,\;\mathrm{pgh}\left( d\right)
=0,\;\deg \left( \gamma \right) =0,  \label{A35a}
\end{equation}
\begin{eqnarray}
&&\sum \alpha _{J}\left( \left[ \chi ^{*\Delta }\right] ,\left[ F\right]
\right) e^{J}\left( \stackrel{(k-1)}{\eta }_{\mu _{1}\cdots \mu
_{k}},\partial _{\left[ \mu _{1}\right. }\stackrel{(k-1)}{\eta }_{\left. \mu
_{2}\cdots \mu _{k+1}\right] }\right)  \nonumber \\
&=&\gamma \left( \mathrm{something}\right) \Leftrightarrow \alpha _{J}=0,\;%
\mathrm{for\;all\;}J,  \label{A36}
\end{eqnarray}
\begin{equation}
d\alpha _{J}\left( \left[ \chi ^{*\Delta }\right] ,\left[ F\right] \right)
=\alpha _{J}^{\prime }\left( \left[ \chi ^{*\Delta }\right] ,\left[ F\right]
\right) ,  \label{A38}
\end{equation}
where 
\begin{equation}
\mathrm{agh}\left( \alpha _{J}^{\prime }\right) =\mathrm{agh}\left( \alpha
_{J}\right) ,\;\deg \left( \alpha _{J}^{\prime }\right) =\deg \left( \alpha
_{J}\right) +1.  \label{A40}
\end{equation}
$\blacksquare $

Theorem \ref{hdinhg} is one of the main tools needed for the computation of $%
H\left( s|d\right) $. In particular, it implies that there is no non-trivial
descent for $H\left( \gamma |d\right) $ in strictly positive antighost
number.

\begin{corollary}
\label{gaplusdb}If $a$ with 
\begin{equation}
\mathrm{agh}\left( a\right) =j>0,\;\mathrm{gh}\left( a\right) =g\geq
-j,\;\deg \left( a\right) =p\leq D,  \label{A50a}
\end{equation}
satisfies the equation 
\begin{equation}
\gamma a+db=0,  \label{A50}
\end{equation}
where 
\begin{equation}
\mathrm{agh}\left( b\right) =j>0,\;\mathrm{gh}\left( b\right) =g+1>-j,\;\deg
\left( b\right) =p-1<D,  \label{A51a}
\end{equation}
then one can always redefine $a$%
\begin{equation}
a\rightarrow a^{\prime }=a+d\nu ,  \label{A51}
\end{equation}
so that 
\begin{equation}
\gamma a^{\prime }=0.  \label{A52}
\end{equation}
\end{corollary}

\emph{Proof} The proof can be done in a standard fashion, like, for
instance, in \cite{hepth04}. Meanwhile, it is worth noticing that the
``current'' $b$ from (\ref{A50}) has the expression 
\begin{equation}
b=-\gamma \nu +df,  \label{A71a}
\end{equation}
with $\gamma f\neq 0$ in general. $\blacksquare $

\subsection{Local cohomology of the Koszul-Tate differential\label{4.3}}

The second essential ingredient in the analysis of the local cohomology $%
H\left( s|d\right) $ is the local cohomology of the Koszul-Tate differential
in pure ghost number zero and in strictly positive antighost numbers, $%
H\left( \delta |d\right) $, also known as the characteristic cohomology. We
recall that the local cohomology $H\left( \delta |d\right) $ is completely
trivial at both strictly positive antighost \textit{and} pure ghost numbers
(for instance, see~\cite{gen1}, Theorem 5.4 and~\cite{commun1}). An element $%
\alpha $ with the properties 
\begin{equation}
\mathrm{agh}\left( \alpha \right) >0,\;\mathrm{pgh}\left( \alpha \right) =0,
\label{propdelta}
\end{equation}
is said to belong to $H\left( \delta |d\right) $ if and only if it is $%
\delta $ closed modulo $d$ 
\begin{equation}
\delta \alpha =dc,\;\mathrm{pgh}\left( c\right) =0.  \label{r80}
\end{equation}
If $\alpha \in H\left( \delta |d\right) $ is a $\delta $-boundary modulo $d$%
\begin{equation}
\alpha =\delta b+dc,\;\mathrm{pgh}\left( \alpha \right) =\mathrm{pgh}\left(
b\right) =\mathrm{pgh}\left( c\right) =0,\;\mathrm{agh}\left( \alpha \right)
=\mathrm{agh}\left( c\right) >0,  \label{r80a}
\end{equation}
we will call it trivial in $H\left( \delta |d\right) $. The solution to the
equation (\ref{r80}) is thus unique up to trivial objects, $\alpha
\rightarrow \alpha +\delta b+dc$. The local cohomology $H\left( \delta
|d\right) $ inherits a natural grading in terms of the antighost number,
such that from now on we will denote by $H_{j}\left( \delta |d\right) $ the
local cohomology of $\delta $ in antighost number $j$. As we have discussed
in Section \ref{2}, the free model under study is a normal gauge theory of
Cauchy order equal to $\left( k+1\right) $. Using the general results from~%
\cite{gen1} (also see~\cite{lingr} and~\cite{multi,gen2}), one can state
that the local cohomology of the Koszul-Tate differential at pure ghost
number zero is trivial in antighost numbers strictly greater than its Cauchy
order 
\begin{equation}
H_{j}\left( \delta |d\right) =0,\;j>k+1.  \label{r81}
\end{equation}

The final tool needed for the calculation of $H\left( s|d\right) $ is the
local cohomology of the Koszul-Tate differential in the space of invariant
polynomials, $H^{\mathrm{inv}}\left( \delta |d\right) $, also called the
invariant characteristic cohomology. It is defined via an equation similar
to (\ref{r80}), but with $\alpha $ and $c$ replaced by invariant
polynomials. Along the same line, the notion of trivial element from $H^{%
\mathrm{inv}}\left( \delta |d\right) $ is revealed by (\ref{r80a}) up to the
precaution that both $b$ and $c$ must be invariant polynomials. It appears
the natural question if the result (\ref{r81}) is still valid in the space
of invariant polynomials. The answer is affirmative 
\begin{equation}
H_{j}^{\mathrm{inv}}\left( \delta |d\right) =0,\;j>k+1  \label{r81c}
\end{equation}
and is proved below, in Theorem \ref{hinvdelta}. Actually, we prove that if $%
\alpha _{j}$ is trivial in $H_{j}\left( \delta |d\right) $, then it can be
taken to be trivial also in $H_{j}^{\mathrm{inv}}\left( \delta |d\right) $.
We consider only the case $j\geq k+1$ since our main scope is to argue the
triviality of $H^{\mathrm{inv}}\left( \delta |d\right) $ in antighost number
strictly greater than $\left( k+1\right) $. First, we prove the following
lemma.

\begin{lemma}
\label{trivinv}Let $\alpha $ be a $\delta $-exact invariant polynomial 
\begin{equation}
\alpha =\delta \beta .  \label{A72}
\end{equation}
Then, $\beta $ can also be taken to be an invariant polynomial.
\end{lemma}

\emph{Proof} Let $v$ be a function of $\left[ \chi ^{*\Delta }\right] $ and $%
\left[ t_{\mu _{1}\cdots \mu _{k}|\nu _{1}\cdots \nu _{k}}\right] $. The
dependence of $v$ on $\left[ t_{\mu _{1}\cdots \mu _{k}|\nu _{1}\cdots \nu
_{k}}\right] $ can be reorganized as a dependence on the curvature and its
derivatives, $\left[ F\right] $, and on 
\begin{equation}
\tilde{t}_{\mu _{1}\cdots \mu _{k}|\nu _{1}\cdots \nu _{k}}=\left\{ t_{\mu
_{1}\cdots \mu _{k}|\nu _{1}\cdots \nu _{k}},\partial t_{\mu _{1}\cdots \mu
_{k}|\nu _{1}\cdots \nu _{k}},\cdots \right\} ,  \label{A72abc}
\end{equation}
where $\tilde{t}_{\mu _{1}\cdots \mu _{k}|\nu _{1}\cdots \nu _{k}}$ are not $%
\gamma $-invariant. If $v$ is $\gamma $-invariant, then it does not involve $%
\tilde{t}_{\mu _{1}\cdots \mu _{k}|\nu _{1}\cdots \nu _{k}}$, i.e., $%
v=\left. v\right| _{\tilde{t}_{\mu _{1}\cdots \mu _{k}|\nu _{1}\cdots \nu
_{k}}=0}$, so we have by hypothesis that 
\begin{equation}
\alpha =\left. \alpha \right| _{\tilde{t}_{\mu _{1}\cdots \mu _{k}|\nu
_{1}\cdots \nu _{k}}=0}.  \label{A73}
\end{equation}
On the other hand, $\beta $ depends in general on $\left[ \chi ^{*\Delta
}\right] $, $\left[ F\right] $ and $\tilde{t}_{\mu _{1}\cdots \mu _{k}|\nu
_{1}\cdots \nu _{k}}$. Making $\tilde{t}_{\mu _{1}\cdots \mu _{k}|\nu
_{1}\cdots \nu _{k}}=0$ in (\ref{A72}), using (\ref{A73}) and taking into
account the fact that $\delta $ commutes with the operation of setting $%
\tilde{t}_{\mu _{1}\cdots \mu _{k}|\nu _{1}\cdots \nu _{k}}$ equal to zero,
we find that 
\begin{equation}
\alpha =\delta \left( \left. \beta \right| _{\tilde{t}_{\mu _{1}\cdots \mu
_{k}|\nu _{1}\cdots \nu _{k}}=0}\right) ,  \label{A74}
\end{equation}
with $\left. \beta \right| _{\tilde{t}_{\mu \nu |\alpha \beta }=0}$
invariant. This proves the lemma. $\blacksquare $

Now, we have the necessary tools for proving the next theorem.

\begin{theorem}
\label{hinvdelta}Let $\alpha _{j}^{p}$ be an invariant polynomial with $\deg
\left( \alpha _{j}^{p}\right) =p$ and $\mathrm{agh}\left( \alpha
_{j}^{p}\right) =j$, which is $\delta $-exact modulo $d$%
\begin{equation}
\alpha _{j}^{p}=\delta \lambda _{j+1}^{p}+d\lambda _{j}^{p-1},\;j\geq k+1.
\label{A75}
\end{equation}
Then, we can choose $\lambda _{j+1}^{p}$ and $\lambda _{j}^{p-1}$ to be
invariant polynomials.
\end{theorem}

\emph{Proof} Initially, by successively acting with $d$ and $\delta $ on (%
\ref{A75}) (see, for instance \cite{hepth04,lingr}) we obtain the tower of
equations 
\[
\alpha _{j+D-p}^{D}=\delta \lambda _{j+D-p+1}^{D}+d\lambda _{j+D-p}^{D-1}, 
\]
\[
\vdots 
\]
\[
\alpha _{j+1}^{p+1}=\delta \lambda _{j+2}^{p+1}+d\lambda _{j+1}^{p}, 
\]
\[
\alpha _{j}^{p}=\delta \lambda _{j+1}^{p}+d\lambda _{j}^{p-1}, 
\]
\[
\alpha _{j-1}^{p-1}=\delta \lambda _{j}^{p-1}+d\lambda _{j-1}^{p-2}, 
\]
\[
\vdots 
\]
\begin{equation}
\alpha _{j-p}^{0}=\delta \lambda _{j-p+1}^{0}\;\mathrm{or}\;\alpha
_{k+1}^{p-j+k+1}=\delta \lambda _{k+2}^{p-j+k+1}+d\lambda _{3}^{p-j+k}.
\label{A80}
\end{equation}
All the $\alpha $'s in the descent (\ref{A80}) are invariant. Using the
general line from \cite{hepth04,lingr} it can be shown that if one of the $%
\lambda $'s in (\ref{A80}) is invariant, then all the other $\lambda $'s can
be taken to be also invariant.

If $j\geq D+k+1$ (and hence $j-p\geq k+1$), the last equation from the
descent (\ref{A80}) for $p=D$ reads as 
\begin{equation}
\alpha _{j-D}^{0}=\delta \lambda _{j-D+1}^{0}.  \label{A83}
\end{equation}
Using Lemma \ref{trivinv}, it results that $\lambda _{j-D+1}^{0}$ can be
taken to be invariant, such that the above arguments lead to the conclusion
that all the $\lambda $'s from the descent can be chosen invariant. As a
consequence, in the first equation from the descent in this situation,
namely, $\alpha _{j}^{D}=\delta \lambda _{j+1}^{D}+d\lambda _{j}^{D-1}$, we
have that both $\lambda _{j+1}^{D}$ and $\lambda _{j}^{D-1}$ are invariant.
Therefore, the theorem is true in form degree $D$ and in all antighost
numbers $j\geq D+k+1$, so it remains to be proved that it holds in form
degree $D$ and in all antighost numbers $k+1\leq j<D+k+1$. This is done
below.

In the sequel we consider the case $p=D$ and $k+1\leq j<D+k+1$. The top
equation from (\ref{A80}), written in dual notations, takes the form 
\begin{equation}
\alpha _{j}=\delta \lambda _{j+1}+\partial _{\mu }\lambda _{j}^{\mu
},\;k+1\leq j<D+k+1.  \label{A84}
\end{equation}
On the other hand, we can express $\alpha _{j}$ in terms of its E.L.
derivatives by means of the homotopy formula 
\begin{eqnarray}
&&\alpha _{j}=\partial _{\mu }c_{j}^{\mu }+\int\nolimits_{0}^{1}d\tau \left(
\sum_{m=0}^{k-1}\frac{\delta ^{R}\alpha _{j}}{\delta \stackrel{(m)}{\eta }%
_{\mu _{1}\cdots \mu _{k}|\nu _{1}\cdots \nu _{k-m-1}}^{*}}\left( \tau
\right) \stackrel{(m)}{\eta }_{\mu _{1}\cdots \mu _{k}|\nu _{1}\cdots \nu
_{k-m-1}}^{*}\right.  \nonumber \\
&&\left. +\frac{\delta ^{R}\alpha _{j}}{\delta t_{\mu _{1}\cdots \mu
_{k}|\nu _{1}\cdots \nu _{k}}^{*}}\left( \tau \right) t_{\mu _{1}\cdots \mu
_{k}|\nu _{1}\cdots \nu _{k}}^{*}+\frac{\delta ^{R}\alpha _{j}}{\delta
t_{\mu _{1}\cdots \mu _{k}|\nu _{1}\cdots \nu _{k}}}\left( \tau \right)
t_{\mu _{1}\cdots \mu _{k}|\nu _{1}\cdots \nu _{k}}\right) ,  \label{A85}
\end{eqnarray}
where 
\begin{equation}
\frac{\delta ^{R}\alpha _{j}}{\delta \stackrel{(m)}{\eta }_{\mu _{1}\cdots
\mu _{k}|\nu _{1}\cdots \nu _{k-m-1}}^{*}}\left( \tau \right) =\frac{\delta
^{R}\alpha _{j}}{\delta \stackrel{(m)}{\eta }_{\mu _{1}\cdots \mu _{k}|\nu
_{1}\cdots \nu _{k-m-1}}^{*}}\left( \tau \left[ t_{\mu _{1}\cdots \mu
_{k}|\nu _{1}\cdots \nu _{k}}\right] ,\tau \left[ \chi ^{*\Delta }\right]
\right)  \label{A85abc}
\end{equation}
and similarly for the other terms. Denoting the E.L. derivatives of $\lambda
_{j+1}$ by 
\begin{equation}
\frac{\delta ^{R}\lambda _{j+1}}{\delta \stackrel{(m)}{\eta }_{\mu
_{1}\cdots \mu _{k}|\nu _{1}\cdots \nu _{k-m-1}}^{*}}=\stackrel{(m)}{G}%
_{j-m-1}^{\mu _{1}\cdots \mu _{k}|\nu _{1}\cdots \nu _{k-m-1}},\;m=\overline{%
0,k-1},  \label{A86}
\end{equation}
\begin{equation}
\frac{\delta ^{R}\lambda _{j+1}}{\delta t_{\mu _{1}\cdots \mu _{k}|\nu
_{1}\cdots \nu _{k}}^{*}}=\stackrel{(k)}{G}_{j}^{\mu _{1}\cdots \mu _{k}|\nu
_{1}\cdots \nu _{k}},  \label{A86abc}
\end{equation}
\begin{equation}
\frac{\delta ^{R}\lambda _{j+1}}{\delta t_{\mu _{1}\cdots \mu _{k}|\nu
_{1}\cdots \nu _{k}}}=L_{j+1}^{\mu _{1}\cdots \mu _{k}|\nu _{1}\cdots \nu
_{k}},  \label{A87}
\end{equation}
and using (\ref{A84}) we find after some computation that the E.L.
derivatives of $\alpha _{j}$ are given by 
\begin{equation}
\frac{\delta ^{R}\alpha _{j}}{\delta \stackrel{(k-1)}{\eta }_{\mu _{1}\cdots
\mu _{k}}^{*}}=\left( -\right) ^{k-1}\delta \stackrel{(k-1)}{G}_{j-k}^{\mu
_{1}\cdots \mu _{k}},  \label{A88}
\end{equation}
\begin{eqnarray}
&&\frac{\delta ^{R}\alpha _{j}}{\delta \stackrel{(m)}{\eta }_{\mu _{1}\cdots
\mu _{k}|\nu _{1}\cdots \nu _{k-m-1}}^{*}}=\left( -\right) ^{m}\left( \delta 
\stackrel{(m)}{G}_{j-m-1}^{\mu _{1}\cdots \mu _{k}|\nu _{1}\cdots \nu
_{k-m-1}}\right.  \nonumber \\
&&-\left( \partial _{{}}^{\left[ \mu _{1}\right. }\stackrel{(m+1)}{G}%
_{j-m-2}^{\left. \mu _{2}\cdots \mu _{k}\right] \left[ \nu _{1}|\nu
_{2}\cdots \nu _{k-m-1}\right] }\right.  \nonumber \\
&&\left. \left. +\left( -\right) ^{k+1}\left( m+2\right) \stackrel{(m+1)}{G}%
_{j-m-2}^{\mu _{1}\cdots \mu _{k}|\left[ \nu _{2}\cdots \nu _{k-m-1},\nu
_{1}\right] }\right) \right) ,\;m=\overline{0,k-2},  \label{A88abc}
\end{eqnarray}
\begin{eqnarray}
\frac{\delta ^{R}\alpha _{j}}{\delta t_{\mu _{1}\cdots \mu _{k}|\nu
_{1}\cdots \nu _{k}}^{*}} &=&-\delta \stackrel{(k)}{G}_{j}^{\mu _{1}\cdots
\mu _{k}|\nu _{1}\cdots \nu _{k}}+\stackrel{(0)}{G}_{j-1}^{\mu _{1}\cdots
\mu _{k}|\left[ \nu _{2}\cdots \nu _{k},\nu _{1}\right] }  \nonumber \\
&&+\stackrel{(0)}{G}_{j-1}^{\nu _{1}\cdots \nu _{k}|\left[ \mu _{2}\cdots
\mu _{k},\mu _{1}\right] },  \label{A89}
\end{eqnarray}
\begin{equation}
\frac{\delta ^{R}\alpha _{j}}{\delta t_{\mu _{1}\cdots \mu _{k}|\nu
_{1}\cdots \nu _{k}}}=\delta L_{j+1}^{\mu _{1}\cdots \mu _{k}|\nu _{1}\cdots
\nu _{k}}-c_{1}\partial _{\mu _{k+1}}\partial _{\nu _{k+1}}\stackrel{(k)}{G}%
_{j}^{\mu _{1}\cdots \mu _{k+1}|\nu _{1}\cdots \nu _{k+1}}.  \label{A90}
\end{equation}
In the above, $\stackrel{(k)}{G}_{j}^{\mu _{1}\cdots \mu _{k+1}|\nu
_{1}\cdots \nu _{k+1}}$ has the same mixed symmetry like the curvature
tensor $F^{\mu _{1}\cdots \mu _{k+1}|\nu _{1}\cdots \nu _{k+1}}$%
\begin{equation}
\stackrel{(k)}{G}_{j\;\;\;\;\;\;\;\;\;\;\;\;\nu _{1}\cdots \nu _{k+1}}^{\mu
_{1}\cdots \mu _{k+1}|}=\sum_{m=0}^{k}\frac{\left( -\right) ^{m}}{m!}\delta
_{\left[ \nu _{1}\right. }^{\left[ \mu _{1}\right. }\cdots \delta _{\nu
_{m}}^{\mu _{m}}\stackrel{(k)}{G}_{j\;\;\;\;\;\;\;\;\;\;\;\;\nu _{m+1}\cdots
\nu _{k}}^{\mu _{m+1}\cdots \mu _{k}|}\delta _{\left. \nu _{k+1}\right]
}^{\left. \mu _{k+1}\right] },  \label{A90a}
\end{equation}
and $\left( \stackrel{(k)}{G}_{j}^{\mu _{m+1}\cdots \mu _{k}|\nu
_{m+1}\cdots \nu _{k}}\right) _{m=\overline{1,k}}$ denote the traces of $%
\stackrel{(k)}{G}_{j}^{\mu _{1}\cdots \mu _{k}|\nu _{1}\cdots \nu _{k}}$
appearing in the right-hand side of the formulas (\ref{A89}--\ref{A90}) 
\begin{equation}
\stackrel{(k)}{G}_{j}^{\mu _{m+1}\cdots \mu _{k}|\nu _{m+1}\cdots \nu
_{k}}=\sigma _{\mu _{1}\nu _{1}}\cdots \sigma _{\mu _{m}\nu _{m}}\stackrel{%
(k)}{G}_{j}^{\mu _{1}\cdots \mu _{k}|\nu _{1}\cdots \nu _{k}},\;m=\overline{%
1,k}.  \label{A90b}
\end{equation}
As the E.L. derivatives of an invariant quantity are also invariant, the
equation in (\ref{A88}) together with Lemma \ref{trivinv} (as $j-k>0$) lead
to 
\begin{equation}
\frac{\delta ^{R}\alpha _{j}}{\delta \stackrel{(k-1)}{\eta }_{\mu _{1}\cdots
\mu _{k}}^{*}}=\left( -\right) ^{k-1}\delta \stackrel{(k-1)}{\bar{G}}%
_{j-k}^{\mu _{1}\cdots \mu _{k}},  \label{A91}
\end{equation}
with $\stackrel{(k-1)}{\bar{G}}_{j-k}^{\mu _{1}\cdots \mu _{k}}$ invariant.
Following a similar reasoning, we find that 
\begin{eqnarray}
&&\frac{\delta ^{R}\alpha _{j}}{\delta \stackrel{(m)}{\eta }_{\mu _{1}\cdots
\mu _{k}|\nu _{1}\cdots \nu _{k-m-1}}^{*}}=\left( -\right) ^{m}\left( \delta 
\stackrel{(m)}{\bar{G}}_{j-m-1}^{\mu _{1}\cdots \mu _{k}|\nu _{1}\cdots \nu
_{k-m-1}}\right.  \nonumber \\
&&-\left( \partial _{{}}^{\left[ \mu _{1}\right. }\stackrel{(m+1)}{\bar{G}}%
_{j-m-2}^{\left. \mu _{2}\cdots \mu _{k}\right] \left[ \nu _{1}|\nu
_{2}\cdots \nu _{k-m-1}\right] }\right.  \nonumber \\
&&\left. \left. +\left( -\right) ^{k+1}\left( m+2\right) \stackrel{(m+1)}{%
\bar{G}}_{j-m-2}^{\mu _{1}\cdots \mu _{k}|\left[ \nu _{2}\cdots \nu
_{k-m-1},\nu _{1}\right] }\right) \right) ,\;m=\overline{0,k-2},  \label{A92}
\end{eqnarray}
\begin{eqnarray}
\frac{\delta ^{R}\alpha _{j}}{\delta t_{\mu _{1}\cdots \mu _{k}|\nu
_{1}\cdots \nu _{k}}^{*}} &=&-\delta \stackrel{(k)}{\bar{G}}_{j}^{\mu
_{1}\cdots \mu _{k}|\nu _{1}\cdots \nu _{k}}+\stackrel{(0)}{\bar{G}}%
_{j-1}^{\mu _{1}\cdots \mu _{k}|\left[ \nu _{2}\cdots \nu _{k},\nu
_{1}\right] }  \nonumber \\
&&+\stackrel{(0)}{\bar{G}}_{j-1}^{\nu _{1}\cdots \nu _{k}|\left[ \mu
_{2}\cdots \mu _{k},\mu _{1}\right] },  \label{A93}
\end{eqnarray}
\begin{equation}
\frac{\delta ^{R}\alpha _{j}}{\delta t_{\mu _{1}\cdots \mu _{k}|\nu
_{1}\cdots \nu _{k}}}=\delta \bar{L}_{j+1}^{\mu _{1}\cdots \mu _{k}|\nu
_{1}\cdots \nu _{k}}-c_{1}\partial _{\mu _{k+1}}\partial _{\nu _{k+1}}%
\stackrel{(k)}{\bar{G}}_{j}^{\mu _{1}\cdots \mu _{k+1}|\nu _{1}\cdots \nu
_{k+1}}.  \label{A94}
\end{equation}
where all the bar quantities are invariant. Since $\alpha _{j}$ is
invariant, it depends on $t_{\mu _{1}\cdots \mu _{k}|\nu _{1}\cdots \nu
_{k}} $ only through the curvature and its derivatives, such that 
\begin{equation}
\frac{\delta ^{R}\alpha _{j}}{\delta t_{\mu _{1}\cdots \mu _{k}|\nu
_{1}\cdots \nu _{k}}}=\partial _{\mu _{k+1}}\partial _{\nu _{k+1}}\Delta
_{j}^{\mu _{1}\cdots \mu _{k+1}|\nu _{1}\cdots \nu _{k+1}},  \label{A95}
\end{equation}
where $\Delta _{j}$ has the mixed symmetry of the curvature tensor. The part
from (\ref{A94}) involving $\stackrel{(k)}{\bar{G}}_{j}$ has a form similar
to that of the right-hand side of (\ref{A95}). Then, $\bar{L}_{j+1}$ must be
expressed in the same manner, i.e., 
\begin{equation}
\delta \bar{L}_{j+1}^{\mu _{1}\cdots \mu _{k}|\nu _{1}\cdots \nu
_{k}}=\partial _{\mu _{k+1}}\partial _{\nu _{k+1}}\Omega _{j}^{\mu
_{1}\cdots \mu _{k+1}|\nu _{1}\cdots \nu _{k+1}},  \label{A96}
\end{equation}
for some $\Omega _{j}$ with the mixed symmetry of the curvature. The
equation (\ref{A96}) shows that for some given indices $\left\{ \nu
_{1}\cdots \nu _{k}\right\} $, the object $\bar{L}_{j+1}^{\mu _{1}\cdots \mu
_{k}|\nu _{1}\cdots \nu _{k}}$ belongs to $H_{j+1}^{D-k}\left( \delta
|d\right) $. As $H_{j+1}^{D-k}\left( \delta |d\right) \simeq
H_{j+2}^{D-k+1}\left( \delta |d\right) \simeq \cdots \simeq
H_{j+k+1}^{D}\left( \delta |d\right) $ (see \cite{gen1}, Theorem 8.1) and $%
H_{j+k+1}^{D}\left( \delta |d\right) \simeq 0$, the equation (\ref{A96})
implies that 
\begin{equation}
\bar{L}_{j+1}^{\mu _{1}\cdots \mu _{k}|\nu _{1}\cdots \nu _{k}}=\delta
R_{j+2}^{\mu _{1}\cdots \mu _{k}||\nu _{1}\cdots \nu _{k}}+\partial _{\rho
}U_{j+1}^{\rho \mu _{1}\cdots \mu _{k}||\nu _{1}\cdots \nu _{k}},
\label{A97}
\end{equation}
where $R_{j+2}$ is separately antisymmetric in the indices $\left\{ \mu
_{1}\cdots \mu _{k}\right\} $ and $\left\{ \nu _{1}\cdots \nu _{k}\right\} $%
, and $U_{j+1}$ is antisymmetric in $\left\{ \rho \mu _{1}\cdots \mu
_{k}\right\} $, as well as in $\left\{ \nu _{1}\cdots \nu _{k}\right\} $.

Now, we prove the theorem in the case $k+1\leq j<D+k+1$ by induction. This
is, we assume that the theorem is valid in antighost number $\left(
j+k+1\right) $ and in form degree $D$, and show that it holds in antighost
number $j$ and in form degree $D$. In agreement with the induction
hypothesis, $R_{j+2}$ and $U_{j+1}$ can be assumed to be invariant. On the
other hand, $\bar{L}_{j+1}^{\mu _{1}\cdots \mu _{k}|\nu _{1}\cdots \nu _{k}}$
must verify the mixed symmetry of the tensor field $t^{\mu _{1}\cdots \mu
_{k}|\nu _{1}\cdots \nu _{k}}$ with respect to the given values $\left\{ \nu
_{1}\cdots \nu _{k}\right\} $, i.e., $\bar{L}_{j+1}^{\mu _{1}\cdots \mu
_{k-1}\left[ \mu _{k}|\nu _{1}\cdots \nu _{k}\right] }=0$, which further
implies that 
\begin{equation}
\delta R_{j+2}^{\mu _{1}\cdots \mu _{k-1}\left[ \mu _{k}||\nu _{1}\cdots \nu
_{k}\right] }+\partial _{\rho }U_{j+1}^{\rho \mu _{1}\cdots \mu _{k-1}\left[
\mu _{k}||\nu _{1}\cdots \nu _{k}\right] }=0.  \label{A98}
\end{equation}
Acting with $\delta $ on (\ref{A98}), we obtain $\partial _{\rho }\left(
\delta U_{j+1}^{\rho \mu _{1}\cdots \mu _{k-1}\left[ \mu _{k}||\nu
_{1}\cdots \nu _{k}\right] }\right) =0$, such that 
\begin{equation}
\delta U_{j+1}^{\rho \mu _{1}\cdots \mu _{k-1}\left[ \mu _{k}||\nu
_{1}\cdots \nu _{k}\right] }=\partial _{\gamma }V_{j}^{\gamma \rho \mu
_{1}\cdots \mu _{k-1}||\mu _{k}\nu _{1}\cdots \nu _{k}},  \label{A99}
\end{equation}
where $V_{j}^{\gamma \rho \mu _{1}\cdots \mu _{k-1}||\mu _{k}\nu _{1}\cdots
\nu _{k}}$ is separately antisymmetric in $\left\{ \gamma \rho \mu
_{1}\cdots \mu _{k-1}\right\} $ and $\left\{ \mu _{k}\nu _{1}\cdots \nu
_{k}\right\} $ (the double bar $||$ signifies that in general this tensor
neither satisfies the identity $V_{j}^{\left[ \gamma \rho \mu _{1}\cdots \mu
_{k-1}||\mu _{k}\right] \nu _{1}\cdots \nu _{k}}\equiv 0$ nor is symmetric
under the inter-change of the two sets of indices). The equation (\ref{A99})
shows that for some fixed indices $\left\{ \mu _{k}\nu _{1}\cdots \nu
_{k}\right\} $, $U_{j+1}^{\rho \mu _{1}\cdots \mu _{k-1}\left[ \mu _{k}||\nu
_{1}\cdots \nu _{k}\right] }$ pertains to $H_{j+1}^{D-k}\left( \delta
|d\right) $, that is finally found isomorphic to $H_{j+k+1}^{D}\left( \delta
|d\right) \simeq 0$, so $U_{j+1}^{\rho \mu _{1}\cdots \mu _{k-1}\left[ \mu
_{k}||\nu _{1}\cdots \nu _{k}\right] }$ is trivial 
\begin{equation}
U_{j+1}^{\rho \mu _{1}\cdots \mu _{k-1}\left[ \mu _{k}||\nu _{1}\cdots \nu
_{k}\right] }=\delta W_{j+2}^{\rho \mu _{1}\cdots \mu _{k-1}\left[ \mu
_{k}||\nu _{1}\cdots \nu _{k}\right] }+\partial _{\gamma }S_{j+1}^{\gamma
\rho \mu _{1}\cdots \mu _{k-1}||\mu _{k}\nu _{1}\cdots \nu _{k}},
\label{A100}
\end{equation}
with $W_{j+2}$ antisymmetric in both $\left\{ \rho \mu _{1}\cdots \mu
_{k-1}\right\} $ and $\left\{ \mu _{k}\nu _{1}\cdots \nu _{k}\right\} $ and $%
S_{j+1}$ separately antisymmetric in $\left\{ \gamma \rho \mu _{1}\cdots \mu
_{k-1}\right\} $ and $\left\{ \mu _{k}\nu _{1}\cdots \nu _{k}\right\} $.
Using again the induction hypothesis, we can assume that $W_{j+2}$ and $%
S_{j+1}$ are invariant. In order to reconstruct $\alpha _{j}$ through the
homotopy formula (\ref{A85}), we need to compute $\delta \bar{L}_{j+1}^{\mu
_{1}\cdots \mu _{k}|\nu _{1}\cdots \nu _{k}}$ by means of formula (\ref{A97}%
), so eventually we need to calculate $\partial _{\rho }U_{j+1}^{\rho \mu
_{1}\cdots \mu _{k-1}\left[ \mu _{k}||\nu _{1}\cdots \nu _{k}\right] }$. In
this respect we use the equation (\ref{A100}) and the identity (that holds
only for a tensor that is separately antisymmetric in its first $\left(
k+1\right) $ and respectively in its last $k$ indices, which does not have
to satisfy any further identity) 
\begin{equation}
U_{j+1}^{\rho \mu _{1}\cdots \mu _{k}||\nu _{1}\cdots \nu _{k}}=\frac{\left(
-\right) ^{k}}{k+1}\sum_{m=0}^{k}\left( \frac{1}{^{k}C_{m}}U_{j+1}^{%
\underline{\underline{\nu _{1}...\nu _{m}}}\underline{\mu _{m+1}...\mu _{k}}%
\left[ \underline{\rho ||\mu _{1}...\mu _{m}}\underline{\underline{\nu
_{m+1}...\nu _{k}}}\right] }\right) ,  \label{A101}
\end{equation}
where two further antisymmetrization should be performed, one over each
underlined group of indices, i.e., $\left\{ \nu _{1}...\nu _{k}\right\} $
and $\left\{ \mu _{1}...\mu _{k}\rho \right\} $. On behalf of the relations (%
\ref{A100}--\ref{A101}), after some computation we obtain that 
\begin{equation}
\partial _{\rho }U_{j+1}^{\rho \mu _{1}\cdots \mu _{k}||\nu _{1}\cdots \nu
_{k}}=\delta \tilde{W}_{j+2}^{\mu _{1}\cdots \mu _{k}||\nu _{1}\cdots \nu
_{k}}+\partial _{\rho }\partial _{\gamma }\tilde{S}_{j+1}^{\mu _{1}\cdots
\mu _{k}\rho ||\nu _{1}\cdots \nu _{k}\gamma },  \label{A102}
\end{equation}
where 
\begin{eqnarray}
&&\tilde{S}_{j+1}^{\mu _{1}\cdots \mu _{k}\rho ||\nu _{1}\cdots \nu
_{k}\gamma }=\frac{\left( -\right) ^{k}}{k+1}\left[ \sum_{m=0}^{\left[ \frac{%
k}{2}\right] }\left( \sum_{i=m}^{\left[ \frac{k}{2}\right] -1}\frac{1}{%
^{k}C_{i}}\left( -\right) ^{i+m}\right. \right.   \nonumber \\
&&\left. +\frac{1}{2^{\varepsilon _{k+1}}}\frac{1}{^{k}C_{\left[ \frac{k}{2}%
\right] }}\left( -\right) ^{m+\left[ \frac{k}{2}\right] }\right) \left(
S_{j+1}^{\underline{\nu _{1}...\nu _{m}}\left[ \mu _{m+1}...\mu _{k}\rho
||\mu _{1}...\mu _{m}\right] \underline{\nu _{m+1}...\nu _{k}\gamma }%
}\right.   \nonumber \\
&&\left. \left. +S_{j+1}^{\underline{\mu _{1}...\mu _{m}}\left[ \nu
_{m+1}...\nu _{k}\gamma ||\nu _{1}...\nu _{m}\right] \underline{\mu
_{m+1}...\mu _{k}\rho }}\right) \right] ,  \label{A103}
\end{eqnarray}
and $\varepsilon _{k+1}$ is defined via 
\begin{equation}
\varepsilon _{k+1}=\left( k+1\right) \;\mathrm{mod}\;2.  \label{noteps}
\end{equation}
On account of (\ref{A103}) it is now obvious that 
\begin{equation}
\tilde{S}_{j+1}^{\mu _{1}\cdots \mu _{k}\rho ||\nu _{1}\cdots \nu _{k}\gamma
}=\tilde{S}_{j+1}^{\mu _{1}\cdots \mu _{k}\rho |\nu _{1}\cdots \nu
_{k}\gamma },  \label{A104b}
\end{equation}
in the sense that it indeed displays the mixed symmetry of the curvature
tensor (it is separately antisymmetric in the indices $\left\{ \mu
_{1}\cdots \mu _{k}\rho \right\} $ and $\left\{ \nu _{1}\cdots \nu
_{k}\gamma \right\} $, but also symmetric under the inter-change of these
two sets of indices, although it does not verify in general the Bianchi I
identity for a $\left( k+1,k+1\right) $ tensor). The tensor $\tilde{S}_{j+1}$
is invariant as $S_{j+1}$ is also invariant. Inserting (\ref{A102}) in (\ref
{A97}) and employing (\ref{A104b}) it results that 
\begin{equation}
\bar{L}_{j+1}^{\mu _{1}\cdots \mu _{k}|\nu _{1}\cdots \nu _{k}}=\delta 
\tilde{R}_{j+2}^{\mu _{1}\cdots \mu _{k}|\nu _{1}\cdots \nu _{k}}+\partial
_{\rho }\partial _{\gamma }\tilde{S}_{j+1}^{\mu _{1}\cdots \mu _{k}\rho |\nu
_{1}\cdots \nu _{k}\gamma }.  \label{A104}
\end{equation}
With the help of (\ref{A91}--\ref{A94}) and (\ref{A104}), the formula (\ref
{A85}) becomes 
\begin{eqnarray}
\alpha _{j} &=&\delta \left[ \int\nolimits_{0}^{1}d\tau \left( \sum_{j=0}^{m}%
\stackrel{(m)}{\bar{G}}_{j-m-1}^{\mu _{1}\cdots \mu _{k}|\nu _{1}\cdots \nu
_{k-m-1}}\stackrel{(m)}{\eta }_{\mu _{1}\cdots \mu _{k}|\nu _{1}\cdots \nu
_{k-m-1}}^{*}\right. \right.   \nonumber \\
&&+\stackrel{(k)}{\bar{G}}_{j}^{\mu _{1}\cdots \mu _{k}|\nu _{1}\cdots \nu
_{k}}t_{\mu _{1}\cdots \mu _{k}|\nu _{1}\cdots \nu _{k}}^{*}  \nonumber \\
&&\left. \left. +\left( \partial _{\rho }\partial _{\gamma }\tilde{S}%
_{j+1}^{\mu _{1}\cdots \mu _{k}\rho |\nu _{1}\cdots \nu _{k}\gamma }\right)
t_{\mu _{1}\cdots \mu _{k}|\nu _{1}\cdots \nu _{k}}\right) \right] +\partial
_{\mu }\sigma _{j}^{\mu }.  \label{A105}
\end{eqnarray}
The last term in the argument of $\delta $ can be written in the form 
\begin{eqnarray}
&&\left( \partial _{\rho }\partial _{\gamma }\tilde{S}_{j+1}^{\mu _{1}\cdots
\mu _{k}\rho |\nu _{1}\cdots \nu _{k}\gamma }\right) t_{\mu _{1}\cdots \mu
_{k}|\nu _{1}\cdots \nu _{k}}=  \nonumber \\
&&\frac{1}{\left( k+1\right) ^{2}}\tilde{S}_{j+1}^{\mu _{1}\cdots \mu
_{k}\rho |\nu _{1}\cdots \nu _{k}\gamma }F_{\mu _{1}\cdots \mu _{k}\rho |\nu
_{1}\cdots \nu _{k}\gamma }+\partial _{\mu }\phi _{j+1}^{\mu },  \label{A106}
\end{eqnarray}
so finally we arrive at 
\begin{eqnarray}
\alpha _{j} &=&\delta \left[ \int\nolimits_{0}^{1}d\tau \left( \sum_{j=0}^{m}%
\stackrel{(m)}{\bar{G}}_{j-m-1}^{\mu _{1}\cdots \mu _{k}|\nu _{1}\cdots \nu
_{k-m-1}}\stackrel{(m)}{\eta }_{\mu _{1}\cdots \mu _{k}|\nu _{1}\cdots \nu
_{k-m-1}}^{*}\right. \right.   \nonumber \\
&&+\stackrel{(k)}{\bar{G}}_{j}^{\mu _{1}\cdots \mu _{k}|\nu _{1}\cdots \nu
_{k}}t_{\mu _{1}\cdots \mu _{k}|\nu _{1}\cdots \nu _{k}}^{*}  \nonumber \\
&&\left. \left. +\frac{1}{\left( k+1\right) ^{2}}\tilde{S}_{j+1}^{\mu
_{1}\cdots \mu _{k}\rho |\nu _{1}\cdots \nu _{k}\gamma }F_{\mu _{1}\cdots
\mu _{k}\rho |\nu _{1}\cdots \nu _{k}\gamma }\right) \right] +\partial _{\mu
}\psi _{j}^{\mu }.  \label{A107}
\end{eqnarray}
We observe that all the terms from the integrand are invariant. In order to
prove that the current $\psi _{j}^{\mu }$ can also be taken invariant, we
switch (\ref{A107}) to the original form notation 
\begin{equation}
\alpha _{j}^{D}=\delta \lambda _{j+1}^{D}+d\lambda _{j}^{D-1},  \label{A107b}
\end{equation}
(where $\lambda _{j}^{D-1}$ is dual to $\psi _{j}^{\mu }$). As $\alpha
_{j}^{D}$ is by assumption invariant and we have shown that $\lambda
_{j+1}^{D}$ can be taken invariant, (\ref{A107b}) becomes 
\begin{equation}
\beta _{j}^{D}=d\lambda _{j}^{D-1}.  \label{A107a}
\end{equation}
It states that the invariant polynomial $\beta _{j}^{D}=\alpha
_{j}^{D}-\delta \lambda _{j+1}^{D}$, of form degree $D$ and of strictly
positive antighost number, is $d$-exact. Then, in agreement with the Theorem 
\ref{hginv} in form degree $D$ (see the paragraph following this theorem),
we can take $\lambda _{j}^{D-1}$ (or, which is the same, $\psi _{j}^{\mu }$)
to be invariant.

In conclusion, the induction hypothesis for antighost number ($j+k+1$) and
form degree $D$ leads to the same property for antighost number $j$ and form
degree $D$, which proves the theorem for all $j\geq k+1$ since we have shown
that it holds for $j\geq D+k+1$. $\blacksquare $

The most important consequence of the last theorem is the validity of the
result (\ref{r81c}) on the triviality of $H^{\mathrm{inv}}\left( \delta
|d\right) $ in antighost number strictly greater than $\left( k+1\right) $.

\subsection{Local cohomology of the BRST differential\label{4.4}}

Now, we have all the necessary tools for the study of the local cohomology $%
H\left( s|d\right) $ in form degree $D$ ($D\geq 2k+1$). We will show that it
is always possible to remove the components of antighost number strictly
greater than $\left( k+1\right) $ from any co-cycle of $H_{D}^{g}\left(
s|d\right) $ in form degree $D$ only by trivial redefinitions.

We consider a co-cycle from $H_{D}^{g}\left( s|d\right) $, $sa+db=0$, with $%
\deg \left( a\right) =D$, $\mathrm{gh}\left( a\right) =g$, $\deg \left(
b\right) =D-1$, $\mathrm{gh}\left( b\right) =g+1$. Trivial redefinitions of $%
a$ and $b$ mean the simultaneous transformations $a\rightarrow a+sc+de$ and $%
b\rightarrow b+df+se$. We expand $a$ and $b$ according to the antighost
number and ask that $a_{0}$ is local, such that each expansion stops at some
finite antighost number \cite{gen2}, $a=\sum\nolimits_{j=0}^{I}a_{j}$, $%
b=\sum\nolimits_{j=0}^{M}b_{j}$, $\mathrm{agh}\left( a_{j}\right) =j=\mathrm{%
agh}\left( b_{j}\right) $. Due to (\ref{k60}), the equation $sa+db=0$ is
equivalent to the tower of equations 
\begin{eqnarray*}
\delta a_{1}+\gamma a_{0}+db_{0} &=&0, \\
&&\vdots \\
\delta a_{I}+\gamma a_{I-1}+db_{I-1} &=&0, \\
&&\vdots
\end{eqnarray*}
The form of the last equation depends on the values of $I$ and $M$, but we
can assume, without loss of generality, that $M=I-1$. Indeed, if $M>I-1$,
the last $\left( M-I\right) $ equations read as $db_{j}=0$, $I<j\leq M$,
which imply that $b_{j}=df_{j}$, $\deg \left( f_{j}\right) =D-2$. We can
thus absorb all the pieces $\left( df_{j}\right) _{I<j\leq M}$ in a trivial
redefinition of $b$, such that the new ``current'' stops at antighost number 
$I$. Accordingly, the bottom equation becomes $\gamma a_{I}+db_{I}=0$, so
the Corollary \ref{gaplusdb} ensures that we can make a redefinition $%
a_{I}\rightarrow a_{I}-d\rho _{I}$ such that $\gamma \left( a_{I}-d\rho
_{I}\right) =0$. Meanwhile, the same corollary (see the formula (\ref{A71a}%
)) leads to $b_{I}=dg_{I}+\gamma \rho _{I}$, where $\deg \left( \rho
_{I}\right) =D-1$, $\deg \left( g_{I}\right) =D-2$, $\mathrm{agh}\left( \rho
_{I}\right) =\mathrm{agh}\left( g_{I}\right) =I$, $\mathrm{gh}\left( \rho
_{I}\right) =g$, $\mathrm{gh}\left( g_{I}\right) =g+1$. Then, it follows
that we can make the trivial redefinitions $a\rightarrow a-d\rho _{I}$ and $%
b\rightarrow b-dg_{I}-s\rho _{I}$, such that the new ``current'' stops at
antighost number $\left( I-1\right) $, while the last component of the
co-cycle from $H_{D}^{g}\left( s|d\right) $ is $\gamma $-closed.

In consequence, we obtained the equation $sa+db=0$, with 
\begin{equation}
a=\sum\limits_{j=0}^{I}a_{j},\;b=\sum\limits_{j=0}^{I-1}b_{j},  \label{exp}
\end{equation}
where $\mathrm{agh}\left( a_{j}\right) =j$ for $0<j<I$ and $\mathrm{agh}%
\left( b_{j}\right) =j$ for $0<j<I-1$. All $a_{j}$ are $D$-forms of ghost
number $g$ and all $b_{j}$ are $\left( D-1\right) $-forms of ghost number $%
\left( g+1\right) $, with $\mathrm{pgh}\left( a_{j}\right) =g+j$ for $0<j<I$
and $\mathrm{pgh}\left( b_{j}\right) =g+j+1$ for $0<j<I-1$. The equation $%
sa+db=0$ is now equivalent with the tower of equations (where some $\left(
b_{j}\right) _{0\leq j\leq I-1}$ could vanish) 
\begin{eqnarray}
\delta a_{1}+\gamma a_{0}+db_{0} &=&0,  \label{A110} \\
&&\vdots  \nonumber \\
\delta a_{j+1}+\gamma a_{j}+db_{j} &=&0,  \label{A110ab} \\
&&\vdots  \nonumber \\
\delta a_{I}+\gamma a_{I-1}+db_{I-1} &=&0,  \label{A109} \\
\gamma a_{I} &=&0.  \label{A108}
\end{eqnarray}
\emph{Next, we show that we can eliminate all the terms }$\left(
a_{j}\right) _{j>k+1}$\emph{\ and }$\left( b_{j}\right) _{j>k}$\emph{\ from
the expansions (\ref{exp}) by trivial redefinitions only.}

Assuming that $a$ stops at a value $L^{\prime }\neq kL$ of the pure ghost
number, $g+I=L^{\prime }$, the bottom equation, (\ref{A108}), yields $%
a_{I}\in H^{L^{\prime }}\left( \gamma \right) $. Then, in agreement with the
result (\ref{k106a}), $a_{I}$ is $\gamma $-trivial, $a_{I}=\gamma \bar{a}%
_{I} $, where $\mathrm{agh}\left( \bar{a}_{I}\right) =I$, $\mathrm{pgh}%
\left( \bar{a}_{I}\right) =g+L^{\prime }-1$ and $\deg \left( \bar{a}%
_{I}\right) =D$. Consequently, we can make the trivial redefinition $%
a\rightarrow a-s\bar{a}_{I}$, whose decomposition stops at antighost number $%
\left( I-1\right) $, such that the bottom equation corresponding to the
redefined co-cycle of $H_{D}^{g}\left( s|d\right) $ takes the form $\gamma
a_{I-1}+db_{I-1}=0$. Now, we apply again the Corollary \ref{gaplusdb} and
replace it with the equation $\gamma a_{I-1}=0$, such that the new
``current'' can be made to end at antighost number $\left( I-2\right) $, $%
b=\sum\limits_{j=0}^{I-2}b_{j} $. In conclusion, if $g+I=L^{\prime }\neq kL$%
, we can always remove the last components $a_{I}$ and $b_{I-1}$ from a
co-cycle $a\in H_{D}^{g}\left( s|d\right) $ and its corresponding
``current'' by trivial redefinitions only.

We can thus assume, without loss of generality, that any co-cycle $a$ from $%
H_{D}^{g}\left( s|d\right) $ can be taken to stop at a value $I$ of the
antighost number such that $g+I=kL$, $a=\sum\limits_{j=0}^{I}a_{j}$, $%
b=\sum\limits_{j=0}^{I-1}b_{j}$. We consider that $I>k+1$. The last equation
from the system equivalent with $sa+db=0$ takes the form (\ref{A108}), with $%
\mathrm{pgh}\left( a_{I}\right) =g+I=kL$, so $a_{I}\in H^{kL}\left( \gamma
\right) $. In agreement with the general results on $H\left( \gamma \right) $
(see Subsection \ref{4.2}) it follows that 
\begin{equation}
a_{I}=\stackrel{(0)}{a}_{I}+\cdots +\stackrel{(L)}{a}_{I}+\gamma \bar{a}_{I},
\label{A110a}
\end{equation}
where 
\begin{equation}
\stackrel{(i)}{a}_{I}=\sum_{J}\alpha _{J,i}e^{J,i},\;i=0,\cdots ,L.
\label{A110b}
\end{equation}
All $\alpha _{J,i}$ are invariant polynomials, with 
\begin{equation}
\mathrm{agh}\left( \alpha _{J,i}\right) =I,\;\deg \left( \alpha
_{J,i}\right) =D,  \label{A110c}
\end{equation}
and 
\begin{eqnarray}
e^{J,i} &\sim &\stackrel{(k-1)}{\eta }_{\stackrel{(1)}{\mu }_{1}\cdots 
\stackrel{(1)}{\mu }_{k}}\cdots \stackrel{(k-1)}{\eta }_{\stackrel{(L-i)}{%
\mu }_{1}\cdots \stackrel{(L-i)}{\mu }_{k}}\times  \nonumber \\
&&\times \partial _{\left[ \alpha _{1}\right. }\stackrel{(k-1)}{\eta }%
_{\left. \stackrel{(L-i+1)}{\mu }_{1}\cdots \stackrel{(L-i+1)}{\mu }%
_{k}\right] }\cdots \partial _{\left[ \alpha _{i}\right. }\stackrel{(k-1)}{%
\eta }_{\left. \stackrel{(L)}{\mu }_{1}\cdots \stackrel{(L)}{\mu }%
_{k}\right] },  \label{eji}
\end{eqnarray}
are the elements of pure ghost number $kL$ of a basis in the space of
polynomials in $\stackrel{(k-1)}{\eta }_{\mu _{1}\cdots \mu _{k}}$ and $%
\partial _{\left[ \mu _{1}\right. }\stackrel{(k-1)}{\eta }_{\left. \mu
_{2}\cdots \mu _{k+1}\right] }$ with the $\bar{D}$-degree equal to $i$.
Applying $\gamma $ on (\ref{A109}) and using (\ref{A108}) together with the
properties (\ref{A35a}) we find that $-d\left( \gamma b_{I-1}\right) =0$,
such that the triviality of the cohomology of $d$ implies that 
\begin{equation}
\gamma b_{I-1}+dc_{I-1}=0,  \label{A111}
\end{equation}
where $\mathrm{agh}\left( c_{I-1}\right) =I-1$, $\mathrm{pgh}\left(
c_{I-1}\right) =kL+1$, $\deg \left( c_{I-1}\right) =D-2$. From the Corollary 
\ref{gaplusdb} it follows (as $I>k+1$ and $k\geq 2$ by assumption, so $I-1>0$%
) that we can make a trivial redefinition such that (\ref{A111}) is replaced
with the equation 
\begin{equation}
\gamma b_{I-1}=0.  \label{A112}
\end{equation}
In agreement with (\ref{A112}), $b_{I-1}$ belongs to $H^{kL}\left( \gamma
\right) $, so we can take 
\begin{equation}
b_{I-1}=\stackrel{(0)}{b}_{I-1}+\cdots +\stackrel{(L)}{b}_{I-1}+\gamma \bar{b%
}_{I-1},  \label{A113}
\end{equation}
where 
\begin{equation}
\stackrel{(i)}{b}_{I-1}=\sum_{J}\beta _{J,i}e^{J,i},\;i=0,\cdots ,L.
\label{A114}
\end{equation}
All $\beta _{J,i}$ are invariant polynomials, with 
\begin{equation}
\mathrm{agh}\left( \beta _{J,i}\right) =I-1,\;\deg \left( \beta
_{J,i}\right) =D-1,  \label{A115}
\end{equation}
and $e^{J,i}$ are the elements (\ref{eji}) with the $\bar{D}$-degree equal
to $i$. Inserting (\ref{A110a}--\ref{A110b}) and (\ref{A113}--\ref{A114}) in
(\ref{A109}) and employing the relation (\ref{eqnew1}) for $b_{I-1}\in
H^{kL}\left( \gamma \right) $, we get that 
\begin{equation}
\sum_{i=0}^{L}\sum_{J}\left[ \pm \left( \delta \alpha _{J,i}+\bar{D}\beta
_{J,i}\right) e^{J,i}+\beta _{J,i}\bar{D}e^{J,i}\right] =\gamma \left(
-a_{I-1}-\hat{b}_{I-1}+\delta \bar{a}_{I}+d\bar{b}_{I-1}\right) ,
\label{A116}
\end{equation}
where $\hat{b}_{I-1}$ comes from $db_{I-1}=\bar{D}b_{I-1}+\gamma \hat{b}%
_{I-1}$. As $\delta \alpha _{J,i}$ and $\bar{D}\beta _{J,i}=d\beta _{J,i}$
are invariant polynomials, while $\bar{D}e^{J,i}=\sum_{J^{\prime
}}A_{J^{\prime },i+1}^{J,i}e^{J^{\prime },i+1}$ (see the formula (\ref{A40H}%
)), the property (\ref{A36}) ensures that the left-hand side of (\ref{A116})
must vanish 
\begin{equation}
\sum_{i=0}^{L}\sum_{J}\left[ \pm \left( \delta \alpha _{J,i}+\bar{D}\beta
_{J,i}\right) e^{J,i}+\beta _{J,i}\bar{D}e^{J,i}\right] =0.  \label{A117}
\end{equation}
Using the decomposition (\ref{A42}) and the definitions (\ref{A43a}--\ref
{A43h}), the projection of the equation (\ref{A117}) on the various values
of the $\bar{D}$-degree becomes equivalent with the equations 
\begin{eqnarray}
0 &:&\delta \alpha _{J,0}+d\beta _{J,0}=0,  \label{A118a} \\
1 &:&\pm \left( \delta \alpha _{J,1}+d\beta _{J,1}\right) +\beta _{J^{\prime
},0}A_{J,1}^{J^{\prime },0}=0,  \label{A118b} \\
&&\vdots  \label{A118c} \\
L &:&\pm \left( \delta \alpha _{J,L}+d\beta _{J,L}\right) +\beta _{J^{\prime
},L-1}A_{J,L}^{J^{\prime },L-1}=0,  \nonumber
\end{eqnarray}
while the equation (\ref{A117}) projected on the value $\left( L+1\right) $
of the $\bar{D}$-degree is automatically satisfied, $\bar{D}_{1}e^{J,L}=0$
due to the relation (\ref{A40C}) and as $e^{J,L}$ contains $L$ factors of
the type $\partial _{\left[ \mu _{1}\right. }\stackrel{(k-1)}{\eta }_{\left.
\mu _{2}\cdots \mu _{k+1}\right] }$.

From (\ref{A118a}) we read that for all $J$ the invariant polynomials $%
\alpha _{J,0}$ belong to $H_{I}^{D}\left( \delta |d\right) $. Thus, as we
assumed that $I>k+1$ and we know that $H_{I}^{D}\left( \delta |d\right) =0$
for $I>k+1$, we deduce that all $\alpha _{J,0}$ are trivial 
\begin{equation}
\alpha _{J,0}=\delta \lambda _{I+1,J,0}^{D}+d\lambda _{I,J,0}^{D-1},
\label{A119}
\end{equation}
where all $\lambda _{I+1,J,0}^{D}$ are $D$-forms of antighost number $\left(
I+1\right) $ and all $\lambda _{I,J,0}^{D-1}$ are $\left( D-1\right) $ forms
of antighost number $I$. Applying the result of the Theorem \ref{hinvdelta},
we have that all $\lambda _{I+1,J,0}^{D}$ and $\lambda _{I+1,J,0}^{D}$ can
be taken to be invariant polynomials, so all $\alpha _{J,0}$ are in fact
trivial in $H_{I}^{\mathrm{inv}D}\left( \delta |d\right) $. Replacing (\ref
{A119}) in (\ref{A118a}) and using $\delta ^{2}=0$ together with $\delta
d+d\delta =0$, we obtain that $d\left( -\delta \lambda _{I,J,0}^{D-1}+\beta
_{J,0}\right) =0$. As $\lambda _{I,J,0}^{D-1}$ and $\beta _{J,0}$ are
invariant polynomials of strictly positive antighost number and of form
degree $\left( D-1\right) $, by Theorem \ref{hginv} it follows that $-\delta
\lambda _{I,J,0}^{D-1}+\beta _{J,0}=d\lambda _{I-1,J,0}^{D-2}$, where $%
\lambda _{I-1,J,0}^{D-2}$ are also invariant polynomials for all $J$, with $%
\mathrm{agh}\left( \lambda _{I-1,J,0}^{D-2}\right) =I-1$ and $\deg \left(
\lambda _{I-1,J,0}^{D-2}\right) =D-2$, so 
\begin{equation}
\beta _{J,0}=\delta \lambda _{I,J,0}^{D-1}+d\lambda _{I-1,J,0}^{D-2}.
\label{A120}
\end{equation}
From (\ref{A119}), we have that 
\begin{eqnarray}
\stackrel{(0)}{a}_{I} &=&\sum_{J}\left( \delta \lambda
_{I+1,J,0}^{D}+d\lambda _{I,J,0}^{D-1}\right) e^{J,0}  \nonumber \\
&=&\pm s\left( \sum_{J}\lambda _{I+1,J,0}^{D}e^{J,0}\right) \pm d\left(
\sum_{J}\lambda _{I,J,0}^{D-1}e^{J,0}\right) \mp \sum_{J}\left( \lambda
_{I,J,0}^{D-1}de^{J,0}\right) .  \label{A121}
\end{eqnarray}
As $de^{J,0}=\sum_{J^{\prime }}A_{J^{\prime },1}^{J,0}e^{J^{\prime
},1}+\gamma \hat{e}^{J,0}$ and $\gamma \lambda _{I,J,0}^{D-1}=0$, we find
that 
\begin{eqnarray}
\stackrel{(0)}{a}_{I} &=&\pm s\left( \sum_{J}\lambda
_{I+1,J,0}^{D}e^{J,0}\right) \pm d\left( \sum_{J}\lambda
_{I,J,0}^{D-1}e^{J,0}\right)  \nonumber \\
&&\mp \gamma \left( \sum_{J}\lambda _{I,J,0}^{D-1}\hat{e}^{J,0}\right) \mp
\sum_{J,J^{\prime }}\left( \lambda _{I,J,0}^{D-1}A_{J^{\prime
},1}^{J,0}e^{J^{\prime },1}\right) .  \label{A122}
\end{eqnarray}
Similarly, relying on (\ref{A120}) we deduce that 
\begin{eqnarray}
\stackrel{(0)}{b}_{I-1} &=&\pm s\left( \sum_{J}\lambda
_{I,J,0}^{D-1}e^{J,0}\right) \pm d\left( \sum_{J}\lambda
_{I-1,J,0}^{D-2}e^{J,0}\right)  \nonumber \\
&&\mp \gamma \left( \sum_{J}\lambda _{I-1,J,0}^{D-2}\hat{e}^{J,0}\right) \mp
\sum_{J,J^{\prime }}\left( \lambda _{I-1,J,0}^{D-2}A_{J^{\prime
},1}^{J,0}e^{J^{\prime },1}\right) .  \label{A123}
\end{eqnarray}
If we perform the trivial redefinitions 
\begin{eqnarray}
a_{I}^{\prime } &=&a_{I}\mp s\left( \sum_{J}\lambda
_{I+1,J,0}^{D}e^{J,0}\right) \mp d\left( \sum_{J}\lambda
_{I,J,0}^{D-1}e^{J,0}\right) ,  \label{A124a} \\
b_{I-1}^{\prime } &=&b_{I-1}\mp s\left( \sum_{J}\lambda
_{I,J,0}^{D-1}e^{J,0}\right) \mp d\left( \sum_{J}\lambda
_{I-1,J,0}^{D-2}e^{J,0}\right) ,  \label{A124b}
\end{eqnarray}
and meanwhile partially fix $\bar{a}_{I}$ and $\bar{b}_{I-1}$ from (\ref
{A110a}) and respectively (\ref{A113}) to 
\begin{eqnarray}
\bar{a}_{I} &=&\pm \sum_{J}\lambda _{I,J,0}^{D-1}\hat{e}^{J,0}+\cdots ,
\label{A124c} \\
\bar{b}_{I-1} &=&\pm \sum_{J}\lambda _{I-1,J,0}^{D-2}\hat{e}^{J,0}+\cdots ,
\label{A124d}
\end{eqnarray}
then (\ref{A122}--\ref{A123}) ensure that the lowest value of the $\bar{D}$%
-degree in the decompositions of $a_{I}^{\prime }$ and $b_{I-1}^{\prime }$
is equal to one. In conclusion, under the hypothesis that $I>k+1$, we
annihilated all the pieces from $a_{I}$ and $b_{I-1}$ with the $\bar{D}$%
-degree equal to zero by trivial redefinitions only. We can then
successively remove the terms of higher $\bar{D}$-degree from $a_{I}$ and $%
b_{I-1}$ by a similar procedure (and also the residual $\gamma $-exact terms
by conveniently fixing the pieces ``$\cdots $'' from $\bar{a}_{I}$ and $\bar{%
b}_{I-1}$) until we completely discard $a_{I}$ and $b_{I-1}$. Next, we pass
to a co-cycle $a$ from $H_{D}^{g}\left( s|d\right) $ that ends at the value $%
\left( I-1\right) $ of the antighost number, and hence $g+I-1\neq kl$, so we
can apply the arguments preceding the equation (\ref{A110a}) and remove both 
$a_{I-1}$ and $b_{I-2}$. This procedure can be continued until we reach
antighost number $\left( k+1\right) $. If $g+k+1$ is $kl$ we cannot go down
and discard $a_{k+1}$ and $b_{k}$, since both $H^{g+k+1}\left( \gamma
\right) $ and $H_{k+1}^{D\mathrm{inv}}\left( \delta |d\right) $ are
non-trivial. However, if $g+k+1\neq kl$, then $H^{g+k+1}\left( \gamma
\right) =0$, so we can go one step lower and remove $a_{k+1}$ and $b_{k}$.
In conclusion, we can take, without loss of generality 
\begin{eqnarray}
a &=&a_{0}+\cdots +a_{k+1},\;b=b_{0}+\cdots +b_{k},\;\mathrm{if}\;g+k+1=kl,
\label{A125} \\
a &=&a_{0}+\cdots +a_{k},\;b=b_{0}+\cdots +b_{k-1},\;\mathrm{if}\;g+k+1\neq
kl,  \label{A126}
\end{eqnarray}
in the equation $sa+db=0$, where $\mathrm{gh}\left( a\right) =g$.
Furthermore, the last terms can be assumed to involve only non-trivial
elements from $H^{\mathrm{inv}}\left( \delta |d\right) $.

\section{Conclusion}

To conclude with, in this paper we have used some specific cohomological
techniques, based on the Lagrangian BRST differential, to show that every
non-trivial co-cycle from the local BRST cohomology in form degree $D$ for a
free, massless tensor field $t_{\mu _{1}\cdots \mu _{k}|\nu _{1}\cdots \nu
_{k}}$ that transforms in an irreducible representation of $GL\left( D,%
\mathbb{R}\right) $, corresponding to a rectangular, two-column Young
diagram with $k>2$ rows, can be taken to stop at antighost number $k$ or $%
\left( k+1\right) $, its last component belonging to $H\left( \gamma \right) 
$ and containing only non-trivial elements from $H^{\mathrm{inv}}\left(
\delta |d\right) $. This result is based on various cohomological properties
involving the exterior longitudinal derivative, the Koszul-Tate
differential, as well as the exterior spacetime differential, which have
been proved in detail. The results contained in this paper are important
from the perspective of constructing consistent interactions that involve
this type of mixed symmetry tensor field since it is known that the
first-order deformation of the solution to the master equation is a co-cycle
of the local BRST cohomology $H_{D}^{0}\left( s|d\right) $ in form degree $D$
and in ghost number zero.

\end{document}